\documentclass[lettersize,journal]{IEEEtran}
\usepackage{amsmath,amsfonts}
\usepackage{array}
\usepackage[table]{xcolor} % Allows coloring in tables
\usepackage{textcomp}
\usepackage{stfloats}
\usepackage{url}
\usepackage{verbatim}
\usepackage{comment}
\usepackage{cite}
\usepackage{float} 
\usepackage{placeins} 
\usepackage{amsmath,amssymb,amsfonts}
\usepackage{graphicx}
\usepackage{textcomp}
\usepackage{xcolor}
\usepackage{hyperref}
\usepackage{caption}
\usepackage{cite}
\usepackage{multirow}
\usepackage{url}
\usepackage{amsmath,amssymb,amsfonts}
\usepackage{graphicx}
\usepackage{textcomp}
\usepackage{caption}
\usepackage{multicol}
\usepackage{lipsum}
\usepackage{xcolor}
\usepackage{subcaption}
\usepackage{array}
\usepackage{float}
\usepackage{algorithm}
\usepackage{textcomp}
\usepackage{scalerel}
\usepackage{cite}
\usepackage{amsmath,amssymb,amsfonts}
\usepackage{algorithmic}
\usepackage{textcomp}
\usepackage{caption}
\usepackage{multicol}
\usepackage{booktabs}   % professional rules
\usepackage{siunitx}    % aligned numbers
\usepackage{xcolor}
\usepackage{esdiff}
\usepackage{cleveref}
\usepackage{etoolbox}
\usepackage{cases}
\usepackage{mathtools}
\usepackage{comment}
\usepackage{amsmath}

\usepackage{tikz}
\usetikzlibrary{shapes.geometric, arrows, positioning}

\allowdisplaybreaks

\usepackage[none]{hyphenat} % Reduce hyphenation
\usepackage{microtype}      % Enhance justification
\begin{document}
	
\title{A Benchmark Library for Distributed Power System Analysis and Optimization}
	
	\author{Milad Hasanzadeh,~\textit{Graduate Student Member, IEEE} and Amin Kargarian, \textit{Senior Member, IEEE}
		\thanks{This work was supported by the National Science Foundation under Grant ECCS-1944752 and Grant ECCS-2312086.
			
			The authors are with the Department of Electrical and Computer Engineering, Louisiana State University, Baton Rouge, LA 70803 USA (email: mhasa42@lsu.edu, kargarian@lsu.edu). DPLib is available at \href{https://github.com/LSU-RAISE-LAB/DPLib.git}{GitHub}. }}

	\maketitle
 
\begin{abstract}
DPLib is an open-source benchmark library created to support research and development in distributed power system analysis and optimization. Unlike centralized tools such as MATPOWER and PGLib, no general purpose, reproducible data library package currently exists for distributed power system studies. DPLib, available at \href{https://github.com/LSU-RAISE-LAB/DPLib.git}{GitHub}, fills this gap by providing 40 multi-region benchmark test cases ranging from 5 buses to 20758 buses, along with a graph-based partitioning toolkit that converts MATPOWER-compatible systems into distributed regional datasets. The toolkit generates standardized \texttt{.mat}, \texttt{.csv}, and \texttt{.m} files, regional MATPOWER version 2 cases, local and global bus mappings, generator and cost assignments, explicit inter-regional tie-line records, and bus-to-region partition maps. It supports unweighted, electrically weighted, and user-defined partitions, and is compared with METIS, KaFFPa, and an IPA-inspired baseline. DPLib also provides ADMM-based distributed DC and AC OPF solvers for validation. Numerical studies report partitioning sensitivity, centralized run times, distributed OPF iterations, run times, and optimality gaps. These results establish DPLib as a reproducible data layer for distributed power system research.
\end{abstract}


\section{Introduction}\label{sec:introduction}

\IEEEPARstart{C}{entralized} frameworks have long served as the dominant approach for modeling and solving large-scale power system problems, including optimal power flow (OPF), planning, control, and unit commitment. A key enabler behind the success and widespread adoption of centralized optimization and analysis in power systems is the availability of standardized benchmark test cases and libraries. Among these, MATPOWER has emerged as a widely accepted benchmark tool, offering standard data for centralized studies~\cite{zimmerman2010matpower,coffrin2018powermodels}. Its role in enabling reproducibility, fair performance comparisons, and methodological development has been fundamental to the evolution of centralized power system analysis.

For centralized studies, for instance, the work in~\cite{lavaei2011zero} proposed convex relaxations for AC OPF and validated the results on MATPOWER test cases. The survey in~\cite{molzahn2019survey} provided a comprehensive review of relaxation techniques for OPF, with all benchmarks sourced from MATPOWER. The work in~\cite{coffrin2015qc} introduced QC relaxation for OPF using MATPOWER data. More recently, the work in~\cite{7552584} developed robust optimization methods for corrective control under uncertainty, with validation using MATPOWER cases. The NESTA archive in~\cite{coffrin2014nesta} enriched MATPOWER with more realistic data while retaining full MATPOWER compatibility. These works, and many more, underscore the importance of a common, reliable data standard facilitated by MATPOWER, which enables researchers to validate and compare centralized algorithms under consistent and reproducible conditions.

In the last decade, distributed and decentralized analysis has gained significant momentum in the power systems community, especially in applications such as distributed OPF, voltage control, state estimation, and demand side management~\cite{molzahn2017survey, kargarian2016toward}. These include studies that apply alternating direction method of multipliers (ADMM) decomposition \cite{kargarian2016toward,erseghe2014distributed}, distributed voltage control \cite{magnusson2020distributed}, parallel and distributed OPF \cite{zhou2011parallel,hasanzadeh2025admm}, consensus-based optimization \cite{peng2015distributed}, and cooperative control frameworks \cite{conte2016distributed,hasanzadeh2024distributed,hasanzadeh2024distributed2}. Additional contributions include scalable distributed frameworks for transmission networks \cite{liu2015consensus}, graph-based coordination strategies \cite{molzahn2017survey}, and distributed control in microgrids \cite{guo2017distributed}. These methods can offer advantages such as improved scalability, privacy preservation, and resilience, depending on the application, communication architecture, and implementation.

However, as the literature on distributed power system methods grows, the supporting data infrastructure has not developed at the same pace. Centralized studies can usually point to a common MATPOWER compatible case and build comparisons around that shared system~\cite{zimmerman2010matpower,babaeinejadsarookolaee2019power}. In contrast, the work in~\cite{muhlpfordt2021distributed} illustrates that a distributed study requires more than the original centralized case. The system must be divided into regions, the local data of each region must be extracted, the inter-regional tie-lines must be identified, and the final regional files must be stored in a format that can be used directly by distributed methods~\cite{hasanzadeh2025admm}. If each paper performs these steps differently, the resulting studies become difficult to reproduce and compare.

This gap is important because distributed power system data are often paper-specific~\cite{molzahn2017survey}. In some studies, regions are defined based on geography, utility areas, electricity markets, or operator boundaries~\cite{guo2016enabling,kekatos2012distributed,vukovic2013security}. In other studies, the regions are created manually or by using graph-based partitioning tools~\cite{chen2015quickest,balasubramaniam2016balanced}. Some datasets are released, but many are not fully documented, not provided in a reusable format, or not compatible with standard power system software. Therefore, when two distributed methods report different results, the difference may stem from the proposed method, the partitioning rule, hidden preprocessing steps, the treatment of boundary buses, or the regional data format. This makes benchmarking in distributed analysis less direct than in centralized analysis.

In response to this need, we have developed DPLib, a standard benchmark library for distributed power system analysis and optimization. DPLib provides 40 multi-region benchmark cases, ranging from 5 to 20758 buses,  generated from MATPOWER-compatible test systems.  The main novelty of DPLib is that it provides a reproducible, ready-to-use, and open-source framework for evaluating distributed power system methods under consistent modeling assumptions, data formats, and regional decomposition structures. In addition to the base distributed benchmark cases, DPLib also includes distributed active power increase (DPLib\_api) variants generated from the corresponding PGLib active-power-increase cases \cite{babaeinejadsarookolaee2019power}. These variants introduce stressed operating conditions, with increased loading and line congestion, enabling the evaluation of distributed solvers in more challenging scenarios.

The first feature of DPLib is to store the generated distributed data in a regional MATPOWER version 2 format \cite{zimmerman2010matpower}. Each generated dataset contains complete regional information, including local buses, local branches, local generators, cost data when available, and a documented table of inter-regional tie-lines. Therefore, DPLib provides complete distributed power system datasets rather than only bus-to-region labels as in \cite{murray2019comparison,li2024gpu}. Beyond distributed OPF, this regional data structure can support other multi-region studies, including multi-region unit commitment, transmission expansion planning, generation expansion planning, voltage stability analysis, distributed security assessment, and reliability screening.

The second feature is to make the partitioning process consistent with the intended use of distributed power system datasets. A power network can be represented as a graph whose nodes are buses and whose edges are transmission lines~\cite{molzahn2017survey,guo2016enabling}; however, DPLib does not treat partitioning merely as an abstract graph-minimization problem. In practice, real power networks often exhibit an inherent regional structure: buses within the same area are typically more strongly connected, while different areas are connected through a smaller number of inter-regional tie-lines~\cite{molzahn2017survey,guo2016enabling,kargarian2016toward,hasanzadeh2025admm}. Such a structure may arise from utility territories, control areas, market zones, geographical boundaries, ownership constraints, or data-management responsibilities. Therefore, useful distributed benchmark datasets should preserve, as much as possible, the natural separation between internally coherent regions and more loosely connected neighboring regions. DPLib uses graph-based partitioning to identify topology-aware regions and selects valid candidates with fewer tie-lines as a practical indicator of limited inter-regional coupling. 

Building on this principle, the third feature is to define partitioning options that preserve the intended meaning of a distributed power system dataset. DPLib supports two such options. The first option is unweighted topological partitioning, which is suitable when the goal is to obtain a topology-driven regional structure with sparse boundary connections. In systems where geographical, market, or ownership boundaries are reflected in network topology, this option may also provide a useful approximation of such regional structure. The second option is electrically weighted partitioning, in which graph edges are weighted using selected line attributes, such as reactance, charging susceptance, and thermal rating. This option allows users to examine how electrical line attributes, in addition to pure connectivity, affect the resulting regional structure. DPLib also supports user-defined partitions when regional boundaries are externally specified.

Prior studies have shown the importance of distributed optimization and algorithmic benchmarking for power system applications, particularly in distributed OPF and related decomposition frameworks~\cite{molzahn2017survey,erseghe2014distributed,guo2015intelligent,alkhraijah2023powermodelsada}. These efforts highlight the need for multi-region test systems, but their primary emphasis is often on algorithm design, convergence behavior, or solver evaluation. In parallel, graph-partitioning methods such as METIS and KaFFPa provide powerful tools for dividing large networks into smaller regions~\cite{karypis1998fast,sanders2011engineering}. However, a reusable benchmark library also requires standardized data construction beyond assigning buses to regions, including regional case files, boundary/tie-line information, consistent indexing, and transparent data formats. DPLib is designed to address this data-layer need.

The fourth feature is to provide both a ready to use benchmark library and a partitioning toolkit. The fixed benchmark cases support direct testing, comparison, and reproducibility, while the toolkit allows users to generate customized multi-region datasets from public or private MATPOWER-compatible systems. To support externally specified regions, DPLib also includes a user-defined mode in which the user provides a two-column bus-to-region matrix, and the toolkit converts that assignment into the same standard regional data format. Thus, DPLib supports direct benchmarking, customized public case generation, and private data conversion within a single framework.

The final feature is to include validation tools that test whether the generated regional files, tie-line records, and boundary mappings can be used together in a complete distributed workflow. DPLib therefore provides distributed DC OPF and AC OPF validation solvers based on ADMM. OPF is used as a representative validation problem because it exercises the main components of the distributed data structure, including regional buses, branches, generators, cost data, tie-lines, and boundary consistency~\cite{molzahn2017survey,erseghe2014distributed,hasanzadeh2025admm}. These solvers are provided as reproducible validation tools for the benchmark data format, rather than as new distributed OPF algorithms. The contributions of DPLib are:
\begin{itemize}
    \item A standardized, open-source benchmark library of 40 multi-region test cases (5 to 20758 buses) derived from MATPOWER-compatible systems.
    \item A graph-based partitioning toolkit that converts centralized MATPOWER cases into complete, standardized multi-region datasets with unweighted, weighted, and user-defined partition options.
    \item ADMM-based distributed DC and AC OPF validation solvers for end-to-end verification of the generated datasets.
\end{itemize}

The remainder of the paper is organized as follows. Section~\ref{sec:modeling} introduces the graph-based modeling of power networks and details the DPLib partitioning toolkit. Section~\ref{sec:codes} describes the validation tools used to test the generated distributed datasets. Section~\ref{sec:case_studies} presents the benchmark case studies generated and tested using DPLib. Finally, Section~\ref{sec:conclusion} concludes the paper.

\section{Graph-based Partitioning and Toolkit Description}\label{sec:modeling}

\begin{figure}[!t]
    \centering
    \captionsetup{font={footnotesize}}
    \includegraphics[width=1\columnwidth]{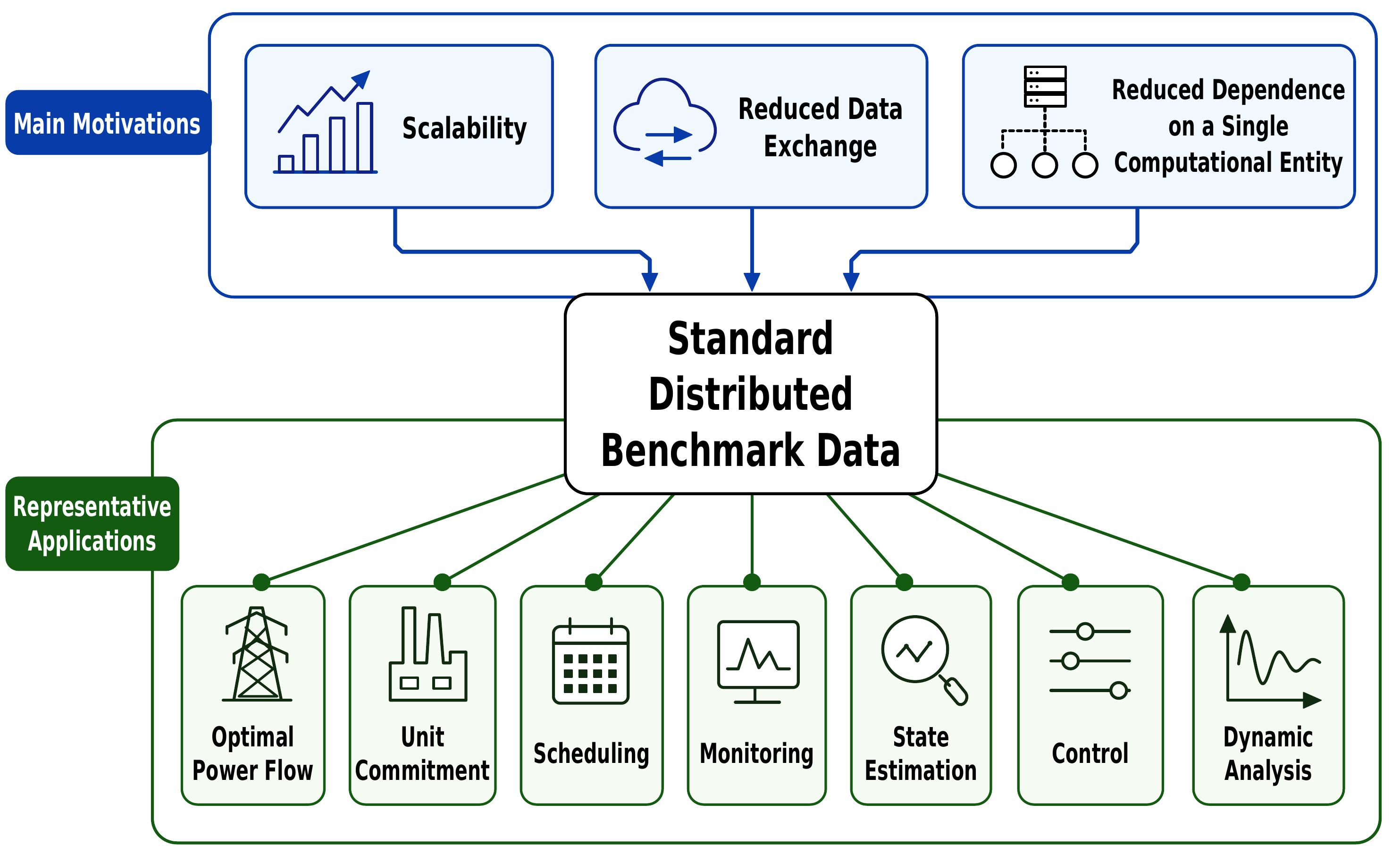}
    \caption{Scope of distributed power system analysis}
    \label{fig:distributed_analysis_scope}
\end{figure}

This section describes the DPLib partitioning toolkit 
and its requirements. All distributed power system applications 
illustrated in Fig.~\ref{fig:distributed_analysis_scope} share a common 
data need, and the following subsections present how DPLib 
constructs this data.

\label{subsec:motivation_scope}
\subsection{Design Requirements for Constructing Distributed Benchmark Data}
Constructing a distributed benchmark library from centralized power system cases requires several design decisions beyond selecting a partitioning algorithm. A centralized MATPOWER case provides one global set of buses, branches, generators, and cost data. A distributed benchmark dataset must convert this single structure into several regional structures while preserving physical consistency, data traceability, and compatibility with existing power system tools. This subsection presents the main design requirements that guided the development of DPLib.

The first requirement is to define a complete and compatible output format for the data conversion process. A direct graph partition produces only a bus-to-region assignment, which is not sufficient for distributed power system studies because the original case must also be decomposed into regional cases, internal and external branches must be separated, generators and cost functions must be mapped to the correct regions, local bus numbering must be created, and inter-regional tie-lines must be stored explicitly. Therefore, DPLib treats a distributed benchmark case as a structured data package composed of regional MATPOWER-compatible cases and a global tie-line table, with each regional file stored in the MATPOWER version 2 structure, i.e., \texttt{mpc.version = '2'}, preserving the familiar \texttt{bus}, \texttt{branch}, \texttt{gen}, and \texttt{gencost} fields when available. The inter-regional tie-lines are stored separately because they define coupling among regions rather than belonging to a single local case, allowing each region to remain a valid local MATPOWER case while retaining the boundary information needed for distributed analysis.

The second requirement is to preserve traceability between the original centralized case and the generated regional cases. During the conversion, some test systems may require bus renumbering, and each region also requires its own local bus numbering. Without a systematic mapping, it becomes difficult to verify whether all original buses, branches, and generators have been preserved. To further support traceability, the toolkit also exports a two-column \texttt{.csv} partition file in which bus number follows the original centralized case numbering rather than the local regional numbering. DPLib therefore performs consistency checks during data generation. It verifies that each original bus is assigned to exactly one region, that all internal branches are included in the corresponding regional cases, that all generators and cost data are mapped to their regions, and that every branch connecting two different regions is recorded as a tie-line. 

The third requirement is to decide how regional boundaries should be generated.If the primary goal were to support a specific numerical decomposition method, a general-purpose graph partitioner could be used to balance region sizes or minimize cut size, though as shown in Table~\ref{tab:partition_method_comparison}, neither criterion alone guarantees faster computation or better convergence. However, a distributed power system benchmark should also support studies that consider regions with physical or operational meaning. For this reason, DPLib does not treat the partitioning step solely as an abstract graph cut problem. In the automatic mode, the network is represented as a graph and partitioned using spectral clustering, grouping buses with strong topological or electrical proximity. The default unweighted formulation uses the physical network topology and avoids case dependent scaling choices. The optional weighted formulation allows electrical parameters to influence the partitioning when the user wants electrical coupling strength reflected in the regional structure.

The fourth requirement is to support both endogenous and exogenous regional definitions. In some benchmark studies, the user may want DPLib to automatically generate regions from the network graph. In other studies, the regions may already be fixed by the application, for example by utility service areas, control areas, market zones, geographical boundaries, ownership boundaries, or institutional jurisdictions. A library limited only to algorithmic partitioning would not support such cases. Therefore, DPLib includes two data generation modes. In the automatic mode, the bus-to-region assignment is produced by the graph-based partitioning procedure. In the user-defined mode, the user provides a two-column matrix of the form \([\text{bus number},\text{region number}]\). DPLib then verifies the assignment and converts it to the same regional MATPOWER-compatible format. Thus, both automatically generated and externally specified partitions lead to the same standardized output structure.

The fifth requirement is to determine whether the library should provide fixed benchmark cases, a data generation tool, or both. Providing only fixed cases would make the library easy to use but would limit users who need different numbers of regions or who want to apply the process to their own systems. Providing only a tool would require every user to generate and validate their own datasets before performing algorithmic studies. DPLib therefore provides both. The ready to use benchmark cases support direct testing and comparison, while the partitioning toolkit supports customized data generation from public or private MATPOWER-compatible systems.

The final requirement is validation. A distributed benchmark dataset should be tested in a workflow that uses the regional files, tie-line table, and boundary information together. DPLib therefore includes distributed DC OPF and AC OPF validation tools that load the generated regional cases, construct boundary variables, coordinate neighboring regions via recorded tie-lines, and monitor convergence.

The following subsection presents the graph model construction used in the automatic partitioning mode.

\subsection{Power Network Modeling and Spectral Clustering}
\label{subsec:graph_spectral_partition}

To prepare the network for partitioning, the power system is modeled as an undirected graph \(
\mathcal{G}=(\mathcal{V},\mathcal{E}),
\) where buses form the nodes and transmission lines form the edges. Let
\(\mathcal{V}=\{1,\ldots,n_b\}\) denote the set of buses and let \(\mathcal{E}\) denote the set of branches. The graph representation is encoded through the adjacency matrix \(A\), and two forms of this matrix may be constructed depending on how connectivity is interpreted.

In the automatic partitioning mode, DPLib seeks a bus-to-region assignment
\[
\pi:\mathcal{V}\rightarrow\{1,\ldots,k\},
\]
where \(k\) is the user-specified number of regions. The set of buses assigned to region \(r\) is
\[
\mathcal{V}_r(\pi)=\{i\in\mathcal{V}\mid \pi_i=r\}, \qquad r=1,\ldots,k.
\]
A feasible regional assignment must assign every bus to exactly one region, i.e.,
\[
\bigcup_{r=1}^{k}\mathcal{V}_r(\pi)=\mathcal{V}, \qquad
\mathcal{V}_r(\pi)\cap\mathcal{V}_s(\pi)=\emptyset,\quad r\neq s.
\]

In its simplest form, the network is viewed purely through its topological structure. Each transmission line indicates the presence of a connection, and the adjacency matrix is
\begin{align}
A_{ij} =
\begin{cases}
1 , & \text{if a line connects buses } i \text{ and } j,\\
0 , & \text{otherwise}.
\end{cases}
\end{align}
This unweighted representation treats all lines uniformly and emphasizes the combinatorial shape of the network. Such a formulation is robust to data irregularities, reflects the way regional boundaries appear in practice when tie-lines are sparse, and preserves the graph-theoretic properties that spectral clustering exploits most directly.

A second option introduces electrical line attributes explicitly. In this case, the weight of an edge is determined from parameters such as line reactance \(x_{ij}\), line charging susceptance \(b_{ij}\), and thermal rating \(S_{\max,ij}\). After normalization, a composite quantity is assigned to each existing edge,
\begin{align}
A_{ij} = w(x_{ij}, b_{ij}, S_{\max,ij}),
\end{align}
so the graph reflects selected electrical attributes in addition to the presence or absence of a connection.

In the current implementation, the weighted adjacency is formed as
\[
A_{ij}=A_{ji}=w_{ij},
\]
where
\[
w_{ij}=|\hat{x}_{ij}|+|\hat{b}_{ij}|+|\hat{S}_{ij}|.
\]
Here, \(\hat{x}_{ij}\), \(\hat{b}_{ij}\), and \(\hat{S}_{ij}\) are normalized versions of line reactance, line charging susceptance, and thermal rating, respectively. Missing, zero, or numerically unusable values are replaced by valid fallback values before constructing the weight. The unweighted mode is the default for benchmark generation, while the weighted mode is provided for users who want selected electrical line attributes to influence the partition.

Although the toolkit supports a weighted Laplacian that incorporates electrical heterogeneity, we treat this option as an auxiliary feature rather than the default mode. Weighted formulations can highlight electrically strong corridors, but doing so requires choices about normalization, scaling, and aggregation of heterogeneous parameters---decisions that are not standardized across power system studies. Because the purpose of DPLib is to produce baseline multi-region datasets for benchmarking distributed algorithms, we use the unweighted formulation to ensure that all results are comparable, repeatable, and free of case dependent parameter tuning. Researchers who wish to emphasize electrical strength rather than topological connectivity may enable the weighted variant, which the toolkit provides for completeness.

Both constructions lead to the degree matrix \(D\) defined by \(D=\mathrm{diag}(A\mathbf{1}),\) and the Laplacian \(L = D - A.\) The Laplacian matrix \(L\) encodes the power network connectivity and topology. For a connected graph, \(L\) is symmetric and positive semi-definite, meaning all its eigenvalues are non-negative. Importantly, the smallest eigenvalue of \(L\) is always zero, corresponding to the trivial uniform eigenvector.

To divide the system into \(k\) regions, DPLib uses spectral clustering, which leverages the properties of \(L\). The first step is to compute the eigenvalues and eigenvectors of \(L\). Among these, the \(k\) eigenvectors associated with the smallest nonzero eigenvalues are selected and used to form the spectral embedding matrix
\[
U=[u_1,\ldots,u_k]\in\mathbb{R}^{n_b\times k}.
\]
Each row \(U_i\) corresponds to bus \(i\) and represents it in a \(k\)-dimensional spectral space, where geometrical proximity reflects topological connectivity.

Before clustering, the rows of \(U\) are normalized as
\[
\bar{U}_i=\frac{U_i}{\|U_i\|_2}, \qquad i=1,\ldots,n_b,
\]
which reduces the effect of magnitude differences in the spectral embedding.

To assign buses into one of \(k\) regions, DPLib applies \(k\)-means clustering to the rows of the spectral embedding. This unsupervised learning algorithm operates by:
\begin{itemize}
    \item Randomly initializing \(k\) centroids;
    \item Assigning each bus to the nearest centroid based on Euclidean distance;
    \item Recalculating the centroid of each cluster;
    \item Repeating the assignment and centroid update steps until convergence.
\end{itemize}

For a given run, this step corresponds to the \(k\)-means problem
\[
\min_{\{c_i\},\{\mu_r\}}
\sum_{i=1}^{n_b}
\left\|\bar{U}_i-\mu_{c_i}\right\|_2^2,
\qquad
c_i\in\{1,\ldots,k\},
\]
where \(c_i\) is the cluster label of bus \(i\), and \(\mu_r\) is the centroid of region \(r\). The resulting labels define a candidate partition \(\pi\). Because \(k\)-means depends on random initialization, DPLib repeats the clustering process several times and evaluates the resulting candidate partitions.

After forming a candidate partition, internal lines and inter-regional tie-lines are identified. For region \(r\), the internal branch set is
\[
\mathcal{E}_r^{\mathrm{int}}(\pi)
=
\{(i,j)\in\mathcal{E}\mid i\in\mathcal{V}_r(\pi),\ j\in\mathcal{V}_r(\pi)\}.
\]
The inter-regional tie-line set is
\[
\mathcal{E}^{\mathrm{tie}}(\pi)
=
\{(i,j)\in\mathcal{E}\mid \pi_i\neq \pi_j\}.
\]
The number of tie-lines is then measured as
\[
J_{\mathrm{tie}}(\pi)=|\mathcal{E}^{\mathrm{tie}}(\pi)|.
\]

This value is used as a candidate-selection score, not as the objective of an exact combinatorial graph partitioning optimization problem. In other words, DPLib does not search over all possible bus partitions. Instead, it generates a finite set of spectral clustering candidates and selects the best valid one.

Each candidate partition is checked before it is accepted. First, every region must contain at least one bus,
\[
|\mathcal{V}_r(\pi)|\geq 1,\qquad r=1,\ldots,k.
\]
Second, for the OPF-based validation workflow used in this paper, every region must contain at least one generator,
\[
|\mathcal{G}_r(\pi)|\geq 1,\qquad r=1,\ldots,k,
\]
where
\[
\mathcal{G}_r(\pi)=\{g\in\mathcal{G}_{\mathrm{gen}}\mid b(g)\in\mathcal{V}_r(\pi)\},
\]
and \(b(g)\) denotes the bus of generator \(g\). Third, the branch accounting condition
\[
\sum_{r=1}^{k}|\mathcal{E}_r^{\mathrm{int}}(\pi)|
+
|\mathcal{E}^{\mathrm{tie}}(\pi)|
=
|\mathcal{E}|
\]
must hold, ensuring that every original branch is represented either as an internal regional branch or as an inter-regional tie-line.

Let \(\Pi_{\mathrm{cand}}\) be the set of candidate partitions generated by repeated spectral clustering runs, and let \(\Pi_{\mathrm{valid}}\subseteq\Pi_{\mathrm{cand}}\) be the subset satisfying the validity checks. The final automatic partition is selected as
\[
\pi^\star
=
\arg\min_{\pi\in\Pi_{\mathrm{valid}}}
J_{\mathrm{tie}}(\pi).
\]
This expression should be interpreted only as a selection rule over the generated candidate set. After \(\pi^\star\) is selected, DPLib constructs the regional MATPOWER compatible cases, locally renumbers buses inside each region, maps generators and cost data to the corresponding regions, identifies the slack bus region, and stores all tie-lines with their complete branch data.

\subsection{DPLib Partitioning Toolkit}

The DPLib partitioning toolkit implements the graph-based clustering framework described in the previous subsection and automates the full workflow for constructing multi-region MATPOWER datasets. The toolkit supports two data generation modes. In automatic mode, the regional partition is generated via spectral clustering using one of the earlier introduced adjacency representations. In the user-defined mode, the user provides an exogenous bus-to-region assignment as a two-column numeric matrix \([\text{bus number},\text{region number}]\), and the toolkit directly converts that assignment into the same DPLib distributed data format. In the automatic mode, the Laplacian is formed using either the default unweighted adjacency or the optional weighted adjacency defined in the previous subsection. The unweighted mode is used for the standard benchmark cases because it depends only on network topology and avoids case-dependent scaling choices. The weighted mode is retained as an optional feature for users who want selected electrical line attributes to affect the generated regional structure.

The workflow proceeds as follows. The case is loaded, and if the original bus numbering is not sequential, all buses, branches, and generators are renumbered from \(1\) to \(n\). In the automatic mode, the adjacency matrix is constructed using either the binary definition or the weighted formulation. The degree matrix and Laplacian are then formed, and the \(k\) Laplacian eigenvectors associated with the smallest nonzero eigenvalues are computed to obtain the spectral embedding. Multiple runs of \(k\)-means are applied to the rows of this matrix to obtain a stable clustering, and the configuration with the smallest number of inter-regional tie-lines is selected. In the user-defined mode, the toolkit instead reads the two-column bus-to-region matrix, verifies that every original bus appears exactly once, checks that region numbers are positive integers, and maps the user provided labels to the internally renumbered bus indices. For the resulting partition in either mode, regional subcases are created by extracting internal buses, branches, generators, and cost data, and each region is locally renumbered to maintain consistency. Tie-lines connecting different regions are identified and stored, along with their full MATPOWER branch data.

\begin{algorithm}[!t]
\caption{DPLib Partitioning Toolkit}\label{alg:partitioning}
\begin{algorithmic}[1]
\STATE Input case name, desired number of regions \( k \), partitioning mode, and optional bus-to-region matrix
\STATE Load data; if bus numbering is non-sequential, renumber buses and update references
\IF{user-defined partitioning mode is selected}
    \STATE Read bus-to-region matrix in the form \([\text{bus number},\text{region number}]\)
    \STATE Verify that every original bus appears exactly once and that all region numbers are positive integers
    \STATE Map the user-defined region labels to the internally renumbered bus indices
    \STATE Set the resulting bus-to-region vector as the fixed partition
\ELSE
    \STATE Construct adjacency matrix \(A\):
    \begin{enumerate}
        \item Unweighted: \(A_{ij}=1\) if a branch connects \(i\) and \(j\)
        \item Weighted: assign \(A_{ij}=w(x_{ij},b_{ij},S_{\max,ij})\)
    \end{enumerate}
    \STATE Form degree matrix \(D\) and Laplacian \(L = D - A\)
    \STATE Compute the Laplacian eigenvectors associated with the \(k\) smallest nonzero eigenvalues and form \(U\)
    \STATE For multiple attempts: apply \(k\)-means to rows of \(U\), build candidate regions, identify tie-lines, and discard candidates without generators
    \STATE Select the valid partition with the minimum number of tie-lines
\ENDIF
\FOR{each region}
    \STATE Extract internal buses, branches, generators, and cost data
    \STATE Locally renumber buses and update regional branch and generator references
\ENDFOR
\STATE Identify the region of the slack bus and record all inter-regional tie-lines
\STATE Export regional cases to \texttt{.mat} and \texttt{.m} files, export the centralized bus-to-region \texttt{.csv} partition map, and generate topology visualization
\end{algorithmic}
\end{algorithm}

The completed partition is exported in several formats. A \texttt{.mat} file contains all regional subcases and a table describing every inter-regional tie-line. A corresponding \texttt{.m} script reproduces the same structures in a clear and readable format. A graphical visualization is produced in \texttt{.png} format, showing regions arranged in a force directed layout together with annotated tie-lines. The toolkit also exports a two-column \texttt{.csv} partition file, \texttt{[bus\_number, region\_number]}. The algorithm for the partitioning toolkit, which illustrates the complete data processing pipeline, is provided in Algorithm \ref{alg:partitioning}.

DPLib also includes a Python interface for reading the generated distributed \texttt{.mat} datasets. This helper allows users to load a DPLib case in Python and store the regional data in dictionary structures, with each region's bus, generator, branch, and cost matrices accessible directly.

\section{Validating DPLib Test Systems Using OPF}\label{sec:codes}
To ensure that the generated multi-region test cases are suitable for distributed power system studies, they must be validated through power system analysis. Although the DPLib partitioning toolkit generates distributed variants of MATPOWER test systems applicable to a wide range of power system studies, including multi-region unit commitment, transmission expansion planning, generation expansion planning, voltage stability analysis, control, state estimation, and security assessment, this section evaluates these test systems for distributed OPF applications only. OPF has been selected as a representative benchmarking problem due to its maturity, availability of centralized references, and compatibility with distributed solvers, supported by over a decade of research in this area. We have developed ADMM-based DC and AC OPF validation solvers that operate directly on the test cases produced by the DPLib partitioning toolkit. These solvers are designed to validate the feasibility and consistency of the partitioned datasets and to demonstrate that the generated regional files can support complete distributed OPF workflows. Nevertheless, users may use alternative solvers or utilize the provided tools to verify and explore their own partitioned datasets.

\begin{figure}[t!]
    \captionsetup{font={footnotesize}}
    \centering
    \begin{subfigure}[b]{0.45\linewidth}
        \centering
        \includegraphics[width=\linewidth]{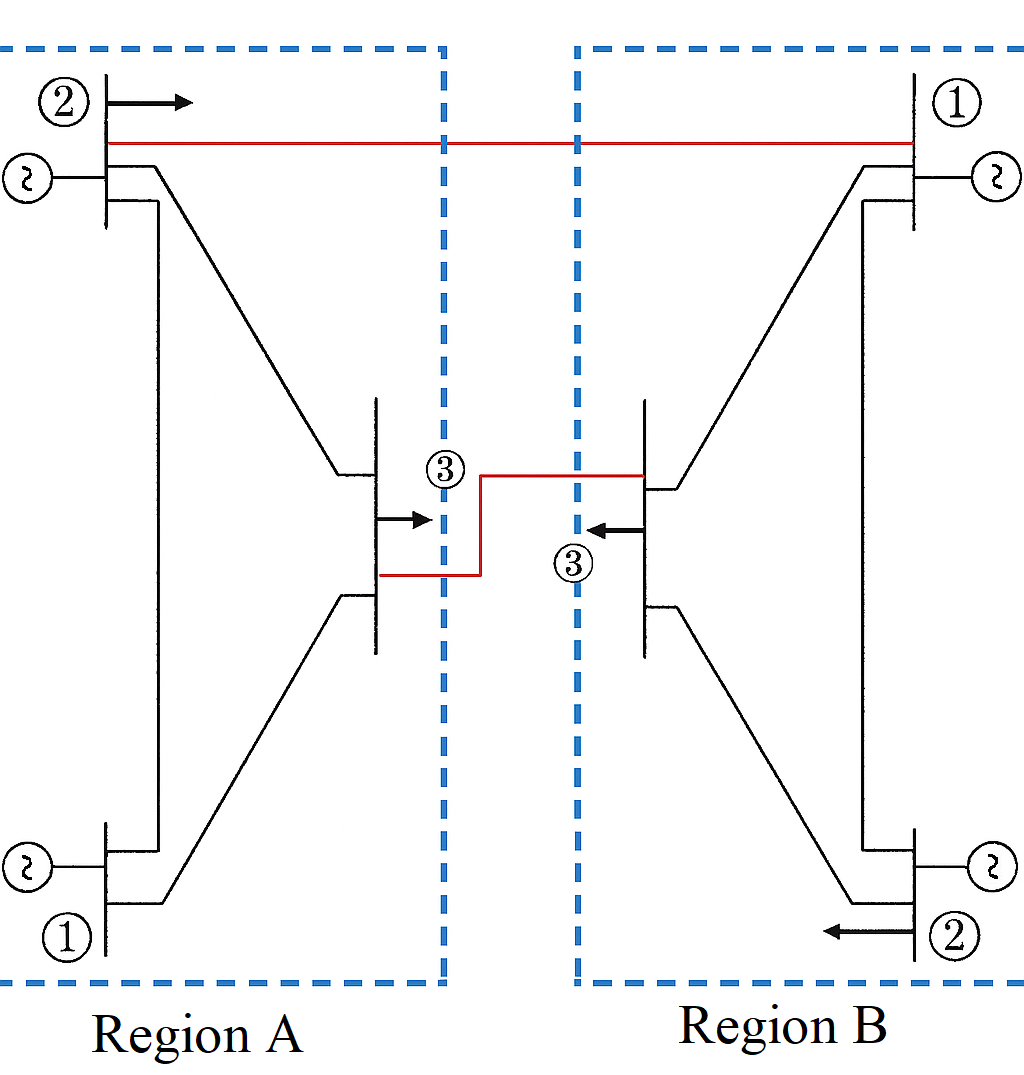}
        \caption{Grid with two regions and red lines are inter-regional transmission tie-lines}
        \label{fig:multi_region}
    \end{subfigure}
    \hfill
    \begin{subfigure}[b]{0.45\linewidth}
        \centering
        \includegraphics[width=\linewidth]{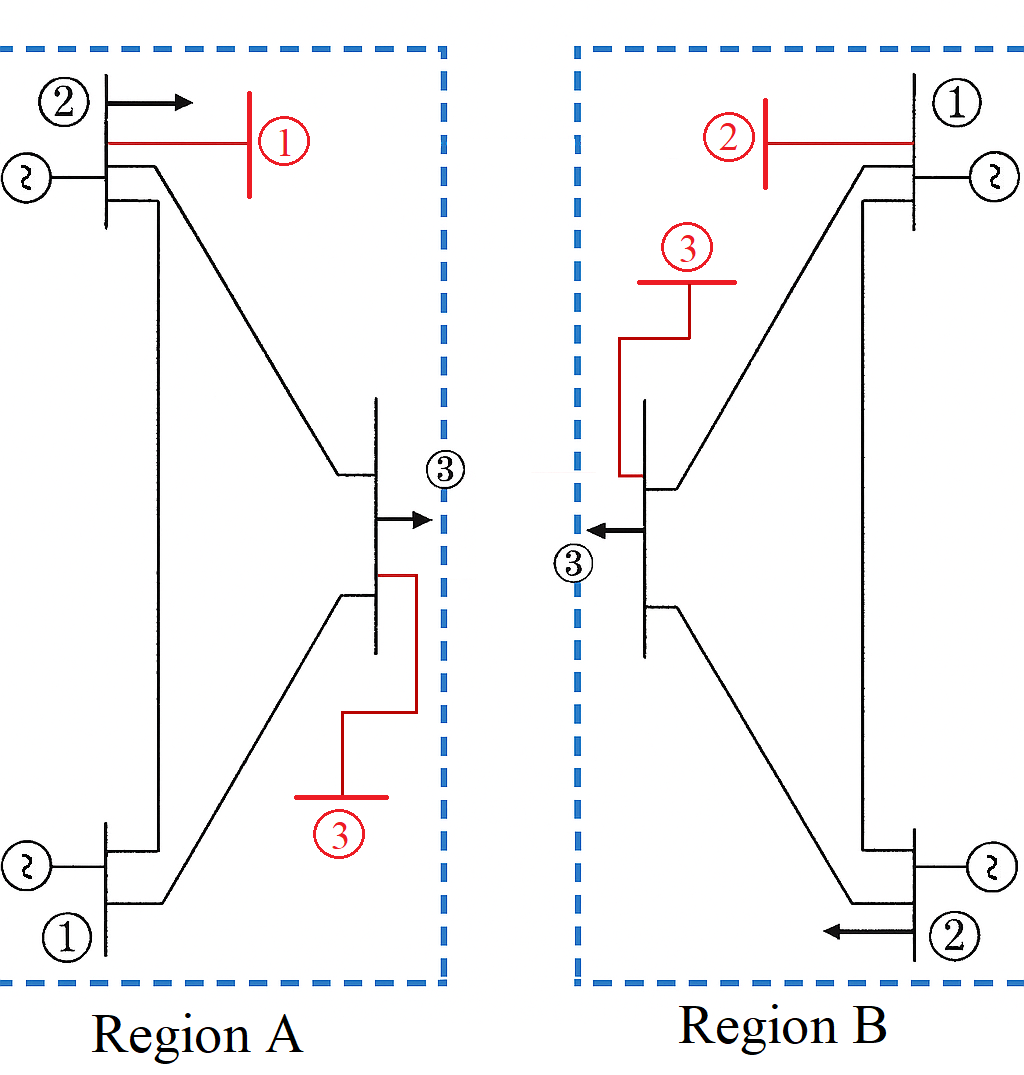}
        \caption{Decomposition into two regions with virtual buses shown in red}
        \label{fig:virtual_buses}
    \end{subfigure}
    \caption{Illustration of a partitioned power system}
    \label{fig:partitioned_vs_virtual}
    \vspace{-9pt}
\end{figure}

\subsection{Distributed OPF Formulation}
\label{sec:opf_formulation}

We formulate the OPF problem in both DC and AC settings for each region \( r \). Each region solves a local OPF problem while maintaining consistency at the boundaries via duplicated (virtual) variables and tie-line constraints \cite{hasanzadeh2025admm}. These copied external bus variables (buses) enable each region to model tie-line interactions locally without requiring direct enforcement of global constraints. The overall structure is illustrated in Fig.~\ref{fig:multi_region}, where a two-region power grid is shown with inter-regional tie-lines in red. Fig.~\ref{fig:virtual_buses} further illustrates how virtual buses (in red) are added to represent shared variables.

\subsubsection{DC OPF Formulation}

The DC OPF subproblem for region \(r\) is written as~\cite{hasanzadeh2025admm}
\begin{subequations}\label{dcopf}
\begin{align}
&\min_{\boldsymbol{x}^{(r)}} 
\sum_{\zeta \in \mathcal{G}_r} f_\zeta\!\left(p_\zeta^{(r)}\right)
\label{dcopf_obj}\\[-1mm]
&\text{s.t.} \nonumber\\
&\underline{p}_\zeta \le p_\zeta^{(r)} \le \bar{p}_\zeta,
\quad \forall \zeta \in \mathcal{G}_r ,
\label{dcopf_gen_limits}\\
&\begin{aligned}
\sum_{\zeta \in \mathcal{G}_r(i)} p_\zeta^{(r)} - p_i^d
&=
\sum_{\substack{j \in \mathcal{B}_r\\ ij \in \mathcal{L}_r^{\mathrm{int}}}}
\frac{\theta_i^{(r)}-\theta_j^{(r)}}{x_{ij}} \\
&\quad+
\sum_{\substack{j \in \mathcal{B}_r^{\mathrm{ext}}\\ ij \in \mathcal{L}_r^{\mathrm{tie}}}}
\frac{\theta_i^{(r)}-\tilde{\theta}_j^{(r)}}{x_{ij}},
\end{aligned}
\quad \forall i \in \mathcal{B}_r ,
\label{dcopf_balance}\\
&\underline{p}_{ij}
\le
\frac{\theta_i^{(r)}-\theta_j^{(r)}}{x_{ij}}
\le
\bar{p}_{ij},
\quad \forall ij \in \mathcal{L}_r^{\mathrm{int}},
\label{dcopf_line_int}\\
&\underline{p}_{ij}
\le
\frac{\theta_i^{(r)}-\tilde{\theta}_j^{(r)}}{x_{ij}}
\le
\bar{p}_{ij},
\quad \forall ij \in \mathcal{L}_r^{\mathrm{tie}},
\label{dcopf_line_tie}\\
&\theta_{i_{\mathrm{ref}}}^{(r)}=0,
\quad i_{\mathrm{ref}}\in\mathcal{B}_r ,
\label{dcopf_ref}\\
&\tilde{\theta}_j^{(r)}=\theta_j^{(s)},
\quad
\begin{aligned}[t]
&\forall ij\in\mathcal{L}_r^{\mathrm{tie}},i\in\mathcal{B}_r,\ j\in\mathcal{B}_s,\ r\ne s,
\end{aligned}
\label{dcopf_cons_j}\\
&\theta_i^{(r)}=\tilde{\theta}_i^{(s)},
\quad
\begin{aligned}[t]
&\forall ij\in\mathcal{L}_r^{\mathrm{tie}},i\in\mathcal{B}_r,\ j\in\mathcal{B}_s,\ r\ne s .
\end{aligned}
\label{dcopf_cons_i}
\end{align}
\end{subequations}

In this formulation, \(r,s\in\mathcal{R}\) are region indices, \(i,j\in\mathcal{B}\) are bus indices, and \(\zeta\in\mathcal{G}\) is a generator index. The notation \(ij\in\mathcal{L}\) denotes the transmission line connecting buses \(i\) and \(j\). The set \(\mathcal{B}_r\subseteq\mathcal{B}\) contains the buses assigned to region \(r\), while \(\mathcal{B}_r^{\mathrm{ext}}\) contains external neighboring buses that are owned by adjacent regions but copied locally in region \(r\) because they are connected to \(\mathcal{B}_r\) through tie-lines. The set \(\mathcal{G}_r\subseteq\mathcal{G}\) contains the generators located in region \(r\), and \(\mathcal{G}_r(i)\subseteq\mathcal{G}_r\) denotes the subset of generators connected to bus \(i\). The set \(\mathcal{L}_r^{\mathrm{int}}\subseteq\mathcal{L}\) contains internal lines whose two endpoints are both in \(\mathcal{B}_r\), while \(\mathcal{L}_r^{\mathrm{tie}}\subseteq\mathcal{L}\) contains tie-lines with one endpoint in \(\mathcal{B}_r\) and the other endpoint in a neighboring region.

The variable \(p_\zeta^{(r)}\) is the active power output of generator \(\zeta\) in region \(r\), and \(f_\zeta(\cdot)\) is its generation cost function. The parameters \(\underline{p}_\zeta\) and \(\bar{p}_\zeta\) are the minimum and maximum active power limits of generator \(\zeta\), respectively. The parameter \(p_i^d\) is the active power demand at bus \(i\), \(x_{ij}\) is the series reactance of line \(ij\), and \(\underline{p}_{ij}\) and \(\bar{p}_{ij}\) are the lower and upper active power flow limits of line \(ij\). The variable \(\theta_i^{(r)}\) is the voltage phase angle of bus \(i\) when bus \(i\) is owned by region \(r\), and \(\tilde{\theta}_j^{(r)}\) is the local copy in region \(r\) of the voltage phase angle of an external bus \(j\in\mathcal{B}_r^{\mathrm{ext}}\). The index \(i_{\mathrm{ref}}\) denotes the global slack bus, 
and constraint~\eqref{dcopf_ref} applies only 
to the region containing it. The decision vector \(\boldsymbol{x}^{(r)}\) collects all variables optimized by region \(r\), including \(p_\zeta^{(r)}\), \(\theta_i^{(r)}\), and \(\tilde{\theta}_j^{(r)}\).

The objective in \eqref{dcopf_obj} minimizes the local generation cost. Constraint \eqref{dcopf_gen_limits} enforces active power generation limits. The nodal balance equation \eqref{dcopf_balance} equates the net active power injection at each owned bus to the sum of DC power flows on internal lines and tie-lines incident to that bus. The first summation in \eqref{dcopf_balance} represents flows on internal lines, while the second summation represents flows on tie-lines through copied external bus angles. Constraints \eqref{dcopf_line_int} and \eqref{dcopf_line_tie} impose active power flow limits on internal lines and tie-lines, respectively. Constraint \eqref{dcopf_ref} fixes the regional reference angle.

The consistency constraints \eqref{dcopf_cons_j} and \eqref{dcopf_cons_i} enforce agreement between owned boundary bus angles and their copied values in neighboring regions. For a tie-line \(ij\) connecting bus \(i\in\mathcal{B}_r\) to bus \(j\in\mathcal{B}_s\), constraint \eqref{dcopf_cons_j} sets the copy of bus \(j\) stored in region \(r\), \(\tilde{\theta}_j^{(r)}\), equal to its physical value \(\theta_j^{(s)}\) in region \(s\). Similarly, \eqref{dcopf_cons_i} sets the copy of bus \(i\) stored in region \(s\), \(\tilde{\theta}_i^{(s)}\), equal to its physical value \(\theta_i^{(r)}\) in region \(r\). 

\subsubsection{AC OPF Formulation}

The AC OPF subproblem for region \(r\) is written as~\cite{hasanzadeh2025admm}
\begin{subequations}\label{acopf}
\begin{align}
    &\min_{\boldsymbol{x}^{(r)}} 
    \sum_{\zeta \in \mathcal{G}_r} f_\zeta\big(p_\zeta^{(r)}\big)
    \label{acopf_obj}\\[1mm]
    &\quad \text{s.t.} \nonumber\\
    &\underline{p}_\zeta \le p_\zeta^{(r)} \le \bar{p}_\zeta,
    \quad \forall \zeta \in \mathcal{G}_r,
    \label{acopf_pg}\\
    &\underline{q}_\zeta \le q_\zeta^{(r)} \le \bar{q}_\zeta,
    \quad \forall \zeta \in \mathcal{G}_r,
    \label{acopf_qg}\\
    &\underline{v}_i \le v_i^{(r)} \le \bar{v}_i,
    \quad \forall i \in \mathcal{B}_r,
    \label{acopf_vlim}\\
    &\begin{aligned}
    \sum_{\zeta \in \mathcal{G}_r(i)} p_\zeta^{(r)}
    - p_i^d
    &
    - g_i^{sh}\big(v_i^{(r)}\big)^2
    =
    \sum_{ij \in \mathcal{L}_{r,f}^{\mathrm{int}}(i)}
    p_{ij,f}^{(r)}
    \\
    &\quad+
    \sum_{ij \in \mathcal{L}_{r,t}^{\mathrm{int}}(i)}
    p_{ij,t}^{(r)} +
    \sum_{\substack{j \in \mathcal{B}_r^{\mathrm{ext}}\\
    ij \in \mathcal{L}_r^{\mathrm{tie}}(i)}}
    \tilde{p}_{ij}^{(r)},
    \end{aligned}
    \quad \forall i \in \mathcal{B}_r,
    \label{acopf_pbal}\\
    &\begin{aligned}
    \sum_{\zeta \in \mathcal{G}_r(i)} q_\zeta^{(r)}
    - q_i^d
    &
    + b_i^{sh}\big(v_i^{(r)}\big)^2=
    \sum_{ij \in \mathcal{L}_{r,f}^{\mathrm{int}}(i)}
    q_{ij,f}^{(r)}
    \\
    &\quad+
    \sum_{ij \in \mathcal{L}_{r,t}^{\mathrm{int}}(i)}
    q_{ij,t}^{(r)} +
    \sum_{\substack{j \in \mathcal{B}_r^{\mathrm{ext}}\\
    ij \in \mathcal{L}_r^{\mathrm{tie}}(i)}}
    \tilde{q}_{ij}^{(r)},
    \end{aligned}
    \quad \forall i \in \mathcal{B}_r,
    \label{acopf_qbal}\\
    &\begin{aligned}
    p_{ij,f}^{(r)} &=
    G_{ij}^{ff}\big(v_i^{(r)}\big)^2
    +G_{ij}^{ft}v_i^{(r)}v_j^{(r)}
    \cos\big(\phi_{ij}^{(r)}\big)\\
    &\quad+
    B_{ij}^{ft}v_i^{(r)}v_j^{(r)}
    \sin\big(\phi_{ij}^{(r)}\big),
    \end{aligned}
    \quad \forall ij \in \mathcal{L}_r^{\mathrm{int}},
    \label{acopf_pff}\\
    &\begin{aligned}
    q_{ij,f}^{(r)} &=
    -B_{ij}^{ff}\big(v_i^{(r)}\big)^2
    -B_{ij}^{ft}v_i^{(r)}v_j^{(r)}
    \cos\big(\phi_{ij}^{(r)}\big)\\
    &\quad+
    G_{ij}^{ft}v_i^{(r)}v_j^{(r)}
    \sin\big(\phi_{ij}^{(r)}\big),
    \end{aligned}
    \quad \forall ij \in \mathcal{L}_r^{\mathrm{int}},
    \label{acopf_qff}\\
    &\begin{aligned}
    p_{ij,t}^{(r)} &=
    G_{ij}^{tt}\big(v_j^{(r)}\big)^2
    +G_{ij}^{tf}v_j^{(r)}v_i^{(r)}
    \cos\big(-\phi_{ij}^{(r)}\big)\\
    &\quad+
    B_{ij}^{tf}v_j^{(r)}v_i^{(r)}
    \sin\big(-\phi_{ij}^{(r)}\big),
    \end{aligned}
    \quad \forall ij \in \mathcal{L}_r^{\mathrm{int}},
    \label{acopf_pft}\\
    &\begin{aligned}
    q_{ij,t}^{(r)} &=
    -B_{ij}^{tt}\big(v_j^{(r)}\big)^2
    -B_{ij}^{tf}v_j^{(r)}v_i^{(r)}
    \cos\big(-\phi_{ij}^{(r)}\big)\\
    &\quad+
    G_{ij}^{tf}v_j^{(r)}v_i^{(r)}
    \sin\big(-\phi_{ij}^{(r)}\big),
    \end{aligned}
    \quad \forall ij \in \mathcal{L}_r^{\mathrm{int}},
    \label{acopf_qft}\\
    &\big(p_{ij,f}^{(r)}\big)^2+\big(q_{ij,f}^{(r)}\big)^2
    \le \bar{s}_{ij}^{2},
    \quad \forall ij \in \mathcal{L}_r^{\mathrm{int}},
    \label{acopf_slimit_f}\\
    &\big(p_{ij,t}^{(r)}\big)^2+\big(q_{ij,t}^{(r)}\big)^2
    \le \bar{s}_{ij}^{2},
    \quad \forall ij \in \mathcal{L}_r^{\mathrm{int}},
    \label{acopf_slimit_t}\\
    &\big(\tilde{p}_{ij}^{(r)}\big)^2+
    \big(\tilde{q}_{ij}^{(r)}\big)^2
    \le \bar{s}_{ij}^{2},
    \quad \forall ij \in \mathcal{L}_r^{\mathrm{tie}},
    \label{acopf_slimit_tie}\\
    &\underline{\Delta\theta}_{ij}
    \le \theta_i^{(r)}-\theta_j^{(r)}
    \le \overline{\Delta\theta}_{ij},
    \quad \forall ij \in \mathcal{L}_r^{\mathrm{int}},
    \label{acopf_dtheta_int}\\
    &\underline{\Delta\theta}_{ij}
    \le \theta_i^{(r)}-\tilde{\theta}_j^{(r)}
    \le \overline{\Delta\theta}_{ij},
    \quad \forall ij \in \mathcal{L}_r^{\mathrm{tie}},
    \label{acopf_dtheta_tie}\\
    &\theta_{i_{\mathrm{ref}}}^{(r)}=0,
    \quad i_{\mathrm{ref}}\in\mathcal{B}_r,
    \label{acopf_ref}\\
    &\tilde{\theta}_j^{(r)}=\theta_j^{(s)},
    \quad
    \begin{aligned}[t]
    &\forall ij\in\mathcal{L}_r^{\mathrm{tie}},
    \ i\in\mathcal{B}_r,\ j\in\mathcal{B}_s,\ r\ne s,
    \end{aligned}
    \label{acopf_cons_theta_j}\\
    &\theta_i^{(r)}=\tilde{\theta}_i^{(s)},
    \quad
    \begin{aligned}[t]
    &\forall ij\in\mathcal{L}_r^{\mathrm{tie}},
    \ i\in\mathcal{B}_r,\ j\in\mathcal{B}_s,\ r\ne s,
    \end{aligned}
    \label{acopf_cons_theta_i}\\
    &\tilde{v}_j^{(r)}=v_j^{(s)},
    \quad
    \begin{aligned}[t]
    &\forall ij\in\mathcal{L}_r^{\mathrm{tie}},
    \ i\in\mathcal{B}_r,\ j\in\mathcal{B}_s,\ r\ne s,
    \end{aligned}
    \label{acopf_cons_v_j}\\
    &v_i^{(r)}=\tilde{v}_i^{(s)},
    \quad
    \begin{aligned}[t]
    &\forall ij\in\mathcal{L}_r^{\mathrm{tie}},
    \ i\in\mathcal{B}_r,\ j\in\mathcal{B}_s,\ r\ne s.
    \end{aligned}
    \label{acopf_cons_v_i}
\end{align}
\end{subequations}

The notation follows the DC OPF formulation in \eqref{dcopf}. The variables
\(p_\zeta^{(r)}\) and \(q_\zeta^{(r)}\) are the active and reactive power
outputs of generator \(\zeta\) in region \(r\), and \(f_\zeta(\cdot)\) is
its generation cost function. The parameters
\(\underline{p}_\zeta,\bar{p}_\zeta\) and
\(\underline{q}_\zeta,\bar{q}_\zeta\) are the active and reactive generation
limits. The parameters \(p_i^d\) and \(q_i^d\) are the active and reactive
demands at bus \(i\). The parameters \(g_i^{sh}\) and \(b_i^{sh}\) denote
the active and reactive shunt coefficients at bus \(i\), respectively. The
variables \(\theta_i^{(r)}\) and \(v_i^{(r)}\) are the voltage angle and
voltage magnitude of bus \(i\) when bus \(i\) is owned by region \(r\), with
\[
\phi_{ij}^{(r)}:=\theta_i^{(r)}-\theta_j^{(r)}.
\]
The variables \(\tilde{\theta}_j^{(r)}\) and \(\tilde{v}_j^{(r)}\) are
local copies in region \(r\) of the voltage angle and voltage magnitude of
an external bus \(j\in\mathcal{B}_r^{\mathrm{ext}}\). The decision vector
\(\boldsymbol{x}^{(r)}\) collects all variables optimized by region \(r\),
including generator outputs, owned bus voltages, and copied external bus
voltage.

For the nodal balance equations, \(\mathcal{L}_{r,f}^{\mathrm{int}}(i)\)
denotes the set of internal lines in region \(r\) for which bus \(i\) is the
from end, and \(\mathcal{L}_{r,t}^{\mathrm{int}}(i)\) denotes the set of
internal lines in region \(r\) for which bus \(i\) is the to end. Therefore,
\(p_{ij,f}^{(r)}\) and \(q_{ij,f}^{(r)}\) are used when bus \(i\) is the
from end of internal line \(ij\), while \(p_{ij,t}^{(r)}\) and
\(q_{ij,t}^{(r)}\) are used when bus \(i\) is the to end of internal line
\(ij\). The same branch index \(ij\) is used for both the from-and
to-end quantities. 

The set \(\mathcal{L}_r^{\mathrm{tie}}(i)\) denotes the set of tie-lines
incident to owned bus \(i\) in region \(r\). The variables
\(\tilde{p}_{ij}^{(r)}\) and \(\tilde{q}_{ij}^{(r)}\) denote the active and
reactive tie-line flows as seen from region \(r\). The variables \(\tilde{p}_{ij}^{(r)}\) and \(\tilde{q}_{ij}^{(r)}\) denote 
the active and reactive tie-line flows as seen from region \(r\). These are computed from \(v_i^{(r)},\theta_i^{(r)}\) and 
the copied variables \(\tilde{v}_j^{(r)},\tilde{\theta}_j^{(r)}\) using the 
same expressions as \eqref{acopf_pff}--\eqref{acopf_qft}. The parameter
\(\bar{s}_{ij}\) is the apparent-power flow limit of line \(ij\). The
parameters \(\underline{\Delta\theta}_{ij}\) and
\(\overline{\Delta\theta}_{ij}\) are the lower and upper voltage
angle-difference limits on line \(ij\), and \(i_{\mathrm{ref}}\) is the global slack bus; constraint~\eqref{acopf_ref} applies only to 
the region containing it.
The branch-admittance coefficients \(G_{ij}^{ff}\), \(G_{ij}^{ft}\),
\(G_{ij}^{tf}\), and \(G_{ij}^{tt}\), and \(B_{ij}^{ff}\), \(B_{ij}^{ft}\),
\(B_{ij}^{tf}\), and \(B_{ij}^{tt}\), are the real and imaginary parts of
the four entries of the \(2\times 2\) branch-admittance matrix of line
\(ij\). The superscripts \(ff\), \(ft\), \(tf\), and \(tt\) denote the
from--from, from--to, to--from, and to--to entries, respectively. These
coefficients are constructed from the standard AC transmission-line
\(\pi\)-model, including series impedance, line charging, and off-nominal
transformer tap ratio. Therefore, \eqref{acopf_pff}--\eqref{acopf_qft} are
the usual polar-coordinate expressions of the branch flows at the from and
to ends.

Constraints \eqref{acopf_pg} and \eqref{acopf_qg} enforce active and
reactive generation limits, while \eqref{acopf_vlim} enforces voltage
magnitude limits. The nodal active- and reactive-power balance equations in
\eqref{acopf_pbal} and \eqref{acopf_qbal} account for generation, demand,
bus shunt terms, the flow contribution of each internal line at the end
incident to the balanced bus, and inter-regional exchanges through
tie-lines via the equivalent injections \(\tilde{p}_{ij}^{(r)}\) and
\(\tilde{q}_{ij}^{(r)}\). Apparent-power limits on internal lines are
enforced by \eqref{acopf_slimit_f} and \eqref{acopf_slimit_t}, while
\eqref{acopf_slimit_tie} limits the tie-line flows as seen from region
\(r\). Angle-difference constraints for internal lines and tie-lines are
given in \eqref{acopf_dtheta_int} and \eqref{acopf_dtheta_tie}, and
\eqref{acopf_ref} fixes the regional reference angle.

Finally, the consistency constraints
\eqref{acopf_cons_theta_j}--\eqref{acopf_cons_v_i} link boundary variables
across neighboring regions. For a tie-line connecting \(i\in\mathcal{B}_r\)
to \(j\in\mathcal{B}_s\), constraints \eqref{acopf_cons_theta_j} and
\eqref{acopf_cons_v_j} enforce that the copied variables
\(\tilde{\theta}_j^{(r)}\) and \(\tilde{v}_j^{(r)}\) of bus \(j\) in region
\(r\) match the owned variables \(\theta_j^{(s)}\) and \(v_j^{(s)}\) in
region \(s\). Similarly, \eqref{acopf_cons_theta_i} and
\eqref{acopf_cons_v_i} enforce that the copied variables
\(\tilde{\theta}_i^{(s)}\) and \(\tilde{v}_i^{(s)}\) of bus \(i\) in region
\(s\) match the owned variables \(\theta_i^{(r)}\) and \(v_i^{(r)}\) in
region \(r\).

\subsubsection{ADMM Decomposition}

The distributed OPF solver uses the sequential ADMM, where regions are updated one at a time in a Gauss–Seidel manner. Each region solves its local OPF problem using the most recent boundary voltage information from neighbors that have already been updated in the current iteration, while using the previous iteration values for neighbors that have not yet been updated. The shared variables are the voltage phase angles and voltage magnitudes of boundary buses. The copied external bus variables introduced in \eqref{dcopf} and \eqref{acopf} represent the neighboring regions' views of these variables.

Before the ADMM iterations start, we also compute problem-dependent scaling factors for the boundary mismatches. For the phase angles, a characteristic scale $\theta_{\text{scale}}$ is obtained from the line reactances and thermal limits as
\[
\theta_{\max,\ell} \approx |x_\ell| \, \frac{\text{rateA}_\ell}{\text{baseMVA}},
\]
where \(x_\ell\) is the reactance of line \(\ell\), \(\text{rateA}_\ell\) is the MATPOWER long-term thermal rating of line \(\ell\), and \(\text{baseMVA}\) is the system power base. Then, $\theta_{\text{scale}}$ is chosen as a robust statistic, for example the median, of $\{\theta_{\max,\ell}\}$ over all lines. For AC OPF, an additional voltage magnitude scale $v_{\text{scale}}$ is defined based on the bus voltage limits, for example, as half the difference between the minimum and maximum allowed magnitudes. These scales are used to normalize the consensus residuals and make them dimensionless.

Consistency across regions is enforced by an augmented Lagrangian of the form
\begin{align}
&L_\rho\left(\theta^{(r)},\tilde{\theta}^{(r)},v^{(r)},\tilde{v}^{(r)},
  \lambda_\theta^{(r)},\lambda_v^{(r)}\right)
  =\sum_{\zeta\in\mathcal{G}_r} f_\zeta\left(p_\zeta^{(r)}\right)
  \nonumber\\&+ \left(\lambda_\theta^{(r)}\right)^\top
  \left(\frac{\theta^{(r)}-\tilde{\theta}^{(r)}}{\theta_{\text{scale}}}\right)
  + \left(\lambda_v^{(r)}\right)^\top
  \left(\frac{v^{(r)}-\tilde{v}^{(r)}}{v_{\text{scale}}}\right)
  \nonumber\\
  &+ \tfrac{\rho_\theta}{2}
  \left\|\frac{\theta^{(r)}-\tilde{\theta}^{(r)}}{\theta_{\text{scale}}}\right\|^2
  + \tfrac{\rho_v}{2}
  \left\|\frac{v^{(r)}-\tilde{v}^{(r)}}{v_{\text{scale}}}\right\|^2 .
\end{align}
which matches the consistency penalized formulation in \cite{hasanzadeh2025admm}. Here  vectors \((\lambda_\theta^{(r)},\lambda_v^{(r)})\) are the multipliers associated with boundary agreement, and \((\rho_\theta,\rho_v)\) are the penalty parameters.

During iteration \(k\), region \(r\) performs the primal update by solving its local AC or DC OPF subproblem. The copied external bus variables from neighboring regions are fixed 
parameters in this optimization. Specifically, during 
the primal update of region \(r\), the owned variables 
\(\theta_i^{(r)}\) and \(v_i^{(r)}\) are optimized, while the copied 
variables \(\tilde{\theta}_j^{(r)}\) and \(\tilde{v}_j^{(r)}\) are 
held fixed at the most recently available values from neighboring 
regions. The augmented Lagrangian pairs each owned boundary variable 
\(\theta_i^{(r)}\) with its copy \(\tilde{\theta}_i^{(s)}\) in region 
\(s\), and each copy \(\tilde{\theta}_j^{(r)}\) with the owned variable 
\(\theta_j^{(s)}\) in region \(s\); the same pairing applies to voltage 
magnitudes in AC OPF. Global consensus is then enforced through the 
dual update steps, which penalize the normalized mismatch between each 
owned variable and its corresponding copy in the neighboring region. After the OPF solve, region \(r\) updates its owned boundary variables \((\theta_i^{(r),k+1},v_i^{(r),k+1})\) and refreshes its copied external variables \((\tilde{\theta}_j^{(r),k+1},\tilde{v}_j^{(r),k+1})\) for tie-lines \(ij\in\mathcal{L}_r^{\mathrm{tie}}\) using the newest available values from neighboring regions. This constitutes the primal update step of ADMM. Once all regions finish their primal updates, the dual variables are updated based on normalized residuals. For each tie-line boundary bus, the updates are
\begin{align}
  \lambda_{\theta}^{(r),k+1}
  &= \lambda_{\theta}^{(r),k}
     + \rho_{\theta}
       \frac{\theta^{(r),k+1}-\tilde{\theta}^{(r),k+1}}
       {\theta_{\text{scale}}}, \\
  \lambda_{v}^{(r),k+1}
  &= \lambda_{v}^{(r),k}
     + \rho_{v}
       \frac{v^{(r),k+1}-\tilde{v}^{(r),k+1}}
       {v_{\text{scale}}}.
\end{align}
In DC OPF, only the angle channel is present, so the update for $\lambda_v$ and the associated terms are omitted. 

Convergence is assessed by the worst-case normalized primal residual across all tie-lines. The primal residual at iteration $k+1$ is computed as
{\small
\[
r_{\mathrm{pri}}^{k+1}
=
\max_{\substack{r\in\mathcal{R}\\ j\in\mathcal{B}_r^{\mathrm{ext}}}}
\left\{
\frac{|\tilde{\theta}_j^{(r),k+1}-\theta_j^{(s),k+1}|}{\theta_{\mathrm{scale}}},
\frac{|\tilde{v}_j^{(r),k+1}-v_j^{(s),k+1}|}{v_{\mathrm{scale}}}
\right\},
\]
}
where \(s\) is the neighboring region that owns the external bus \(j\). This matches the residual definition used in \cite{hasanzadeh2025admm}, up to the normalization by \((\theta_{\text{scale}}, v_{\text{scale}})\).

To improve robustness across different network topologies and operating points, the penalty parameters are updated adaptively based on the relative sizes of the primal and dual residuals, following the heuristic in~\cite{neal2011distributed}. Here, \(r_{\text{dual}}^k\) denotes the dual residual, 
computed as the infinity-norm of the change in the normalized consensus residuals between two consecutive ADMM iterations, consistent with the residual balancing heuristic in~\cite{neal2011distributed}. For the distributed DC OPF solver, a single parameter is updated according to
\[
\rho^{k+1}
=
\begin{cases}
\tau_{\text{incr}} \, \rho^{k}, & \text{if } r_{\text{pri}}^{k} > \mu \, r_{\text{dual}}^{k}, \\[0.2em]
\rho^{k} / \tau_{\text{decr}},   & \text{if } r_{\text{dual}}^{k} > \mu \, r_{\text{pri}}^{k}, \\[0.2em]
\rho^{k},                        & \text{otherwise},
\end{cases}
\]
with chosen parameters $\mu$, $\tau_{\text{incr}}$, and $\tau_{\text{decr}}$. For the distributed AC OPF solver, the same heuristic is applied separately to the angle and magnitude channels, i.e., to $\rho_\theta$ and $\rho_v$, using different balance factors for the $\theta$ and $v$ residuals.

To prevent numerical blow-up and avoid instability caused by excessively large penalties, we enforce an upper bound on the adaptive updates. Specifically, the penalty updates are terminated once the penalty reaches a prescribed maximum value: $\rho_{\mathrm{DC}}^{\max}$ for the distributed DC OPF solver and $\rho_{\mathrm{AC}}^{\max}$ for the distributed AC OPF solver. In practice, if an update would increase the current penalty beyond the corresponding cap, we set $\rho \leftarrow \rho_{\mathrm{DC}}^{\max}$ (DC) or $\rho_\theta,\rho_v \leftarrow \rho_{\mathrm{AC}}^{\max}$ (AC) and disable further penalty adaptation thereafter. This safeguard ensures the algorithm does not diverge due to runaway growth of $\rho$.

Moreover, to avoid overly frequent parameter changes---which can introduce oscillations and hinder convergence---we apply a cooldown period after each penalty update. Concretely, once $\rho$ (DC) or $\rho_\theta,\rho_v$ (AC) is updated, we postpone any subsequent residual balance checks and penalty updates for the next $N_{\mathrm{wait}}$ iterations, where $N_{\mathrm{wait}} = 20$ for cases with fewer than $2000$ buses and $N_{\mathrm{wait}} = 40$ for cases with $2000$ buses or more. This heuristic empirically stabilizes the convergence behavior by allowing the primal and dual iterates to settle under the new penalty before further adaptation is attempted.

Overall, this adaptive strategy automatically increases the penalty when the primal residual dominates and decreases it when the dual residual dominates, promoting a better balance between consensus enforcement and dual convergence while remaining numerically stable through the maximum penalty cap and the update cooldown mechanism.

\subsection{ADMM-Based OPF Validation Solvers}
\label{subsec:admm_solvers}

We have developed two open-source validation solvers for distributed OPF problems, one for DC OPF and another for AC OPF. Both solvers use ADMM to coordinate regional solutions and ensure consistency across tie-lines. 

\subsubsection{DPLib DC OPF Solver}
\label{subsec:dcopf}

The DPLib DC OPF module implements a distributed DC OPF solver based on the ADMM. The MATLAB implementation, available in the file DistDCOPF.m and publicly accessible on \href{https://github.com/LSU-RAISE-LAB/DPLib/tree/main/DPLib%20toolkit/DPLib_DCOPF_Solver}{GitHub}, is designed for partitioned MATPOWER cases and supports the YALMIP optimization backends in MATLAB. The solver requires MATLAB R2020a or newer with MATPOWER installed; YALMIP must also be added to the MATLAB path.

At execution time, the user provides the partitioned MATPOWER file, the ADMM penalty parameter rho, a convergence tolerance, a maximum number of iterations, a $\rho_{\mathrm{DC}}^{\max}$, and optionally the centralized OPF cost for computing the optimality gap. YALMIP emphasizes modeling clarity and solver flexibility, and is suitable for rapid debugging.

After receiving user inputs, DistDCOPF.m loads all regional data, reconstructs inter-regional tie-lines, and augments each region's branch matrix accordingly. Bus angle variables, tie-line indices, and interface bus mappings are then initialized, along with the ADMM dual variables. During each ADMM iteration, every region solves its local DC OPF problem independently using the selected backend. The local objective minimizes the generation cost subject to regional constraints and tie-line consistency penalties. Following all local solves, the solver computes the angle mismatches on shared tie-lines, updates the dual variables, records the maximum primal residual, and checks the convergence condition. If a centralized optimal cost is available, the optimality gap is also computed and plotted. The procedure continues until convergence is reached or the iteration limit is met.

To facilitate reading of the manuscript and implementation of the DPLib codes, we also identify the main code-level names used by the OPF solvers. The regional cases \(\mathcal{B}_r\), \(\mathcal{G}_r\), and \(\mathcal{L}_r\) are stored in \texttt{mpc\_regionRr.bus}, \texttt{mpc\_regionRr.gen}, and \texttt{mpc\_regionRr.branch}, respectively, while inter-regional tie-lines are stored in \texttt{inter-regional\_tielines\_total} and \texttt{inter-regional\_tielinesRr}. In the DC solver, the boundary angle variables \(\theta_i^{(r)}\) and \(\tilde{\theta}_j^{(r)}\) are stored in \texttt{phase\_angles\_Rr}, the angle dual variables are stored in \texttt{lambda}, and the ADMM penalty is \texttt{rho}. In the AC solver, the boundary angle and voltage magnitude variables are stored in \texttt{phase\_angles\_Rr} and \texttt{voltage\_mag\_Rr}, the corresponding dual variables are \texttt{landa} and \texttt{lambdav}, and the penalty parameters are \texttt{rho\_theta} and \texttt{rho\_V}. The residual and performance histories are stored in \texttt{errorLog}, \texttt{dualLog}, and \texttt{optimalityGap}, while the convergence tolerance, maximum iteration count, and maximum penalty cap are passed as \texttt{residualThreshold}, \texttt{maxIterations}, and \texttt{holds}.

The solver supports any number of regions and enables reproducible distributed OPF studies. Its overall workflow is summarized in Algorithm~\ref{alg:opf}.

\begin{algorithm}[t!]
\caption{DPLib Distributed OPF Validation Solver}
\label{alg:opf}
\begin{algorithmic}[1]
    \STATE \textbf{Input:} \texttt{partitionedDataFile}, \texttt{rho}, \texttt{residualThreshold}, \texttt{maxIterations}, \texttt{centralizedCost}, and penalty cap \texttt{holds}
    \STATE Load \texttt{num\_regions}, \texttt{mpc\_regionRr}, and \texttt{inter-regional\_tielines\_total}
    \FOR{each region \(r\)}
        \STATE Build \texttt{inter-regional\_tielinesRr} and augment \texttt{mpc\_regionRr.branch} with local tie-line rows
        \STATE Initialize \texttt{nbRr}, \texttt{nitRr}, \texttt{conRr}, \texttt{YonRr}, and boundary bus map \texttt{region\_Rr}
        \STATE Initialize \texttt{phase\_angles\_Rr}; for AC also initialize \texttt{voltage\_mag\_Rr}
    \ENDFOR
    \STATE Compute scaling factors \texttt{theta\_scale} and, for AC, \texttt{V\_scale}
    \STATE Initialize dual variables: \texttt{lambda} for DC, or \texttt{landa} and \texttt{lambdav} for AC
    \FOR{\(k=1,\ldots,\texttt{maxIterations}\)}
        \FOR{each region \(r\)}
            \IF{DC validation}
                \STATE Solve local DC OPF using \texttt{DCOPF\_SP.m} and update \texttt{phase\_angles\_Rr}
            \ELSE
                \STATE Solve local AC OPF using \texttt{acopf(reg)} and update \texttt{phase\_angles\_Rr} and \texttt{voltage\_mag\_Rr}
            \ENDIF
            \STATE Add regional generation cost to \texttt{currentObjective}
        \ENDFOR
        \STATE Compute normalized boundary mismatches across \texttt{inter-regional\_tielines\_total}
        \STATE Update \texttt{lambda} for DC, or \texttt{landa} and \texttt{lambdav} for AC
        \STATE Store worst primal residual in \texttt{errorLog} and compute dual residual in \texttt{dualLog}
        \STATE If \texttt{centralizedCost} is available, update \texttt{optimalityGap}
        \STATE Update \texttt{rho}, or \texttt{rho\_theta} and \texttt{rho\_V}, using residual balancing, cooldown, and cap \texttt{holds}
        \IF{\(\texttt{errorLog}(k)\le \texttt{residualThreshold}\)}
            \STATE Stop and return convergence results
        \ENDIF
    \ENDFOR
    \STATE Return iteration count, residual history, optimality gap, final objective, dual variables, and tie-line data
\end{algorithmic}
\end{algorithm}

\subsubsection{DPLib AC OPF Solver}
\label{sec:acopf_code}

The DPLib AC OPF solver provides a MATLAB implementation of a distributed ADMM-based AC OPF algorithm that operates on any power system previously partitioned using the DPLib framework. The solver is implemented in the publicly available file \texttt{DistACOPF.m}, accessible through the GitHub repository at \href{https://github.com/LSU-RAISE-LAB/DPLib/tree/main/DPLib%20toolkit/DPLib_ACOPF_Solver}{GitHub}. The code requires a properly configured installation of MATPOWER and IPOPT.

The function begins by requesting the partitioned \texttt{.mat} file and the user-defined ADMM parameters, including the penalty value \(\rho\), the convergence tolerance, the maximum number of iterations, the $\rho_{\mathrm{AC}}^{\max}$, and optionally the centralized AC OPF cost for computing the optimality gap. After these inputs are provided, the solver loads the partitioned data and augments the branch matrices for each region with their associated tie-lines. It then initializes regional variables describing bus counts, voltage magnitudes and angles, and other quantities needed to enable the local AC OPF solves and the coordination steps.

The main ADMM loop is implemented in the \texttt{DistACOPF.m} function. In each iteration, every region independently solves its local nonlinear AC OPF using the helper function \texttt{acopf}, which is based on the formulation in equation~\eqref{acopf}. The updated voltage magnitudes and phase angles from each region are then exchanged virtually via the consistency constraints applied to the tie-lines. The dual variables are updated at each step based on mismatches in the angle and magnitude variables across the interfaces, ensuring that neighboring regions gradually reach agreement. The loop continues until the maximum primal residual across all tie-lines drops below the specified tolerance. If a centralized AC OPF cost is provided, the function also evaluates the optimality gap at every iteration to quantify the progress toward the optimum.

Once the iterations terminate, the solver prints summary information and generates plots that illustrate the convergence behavior of the primal residual and, when applicable, the optimality gap. The complete algorithmic flow implemented in \texttt{DistACOPF.m} is summarized in Algorithm~\ref{alg:opf}.

DPLib provides an automatic main runner for benchmark generation and validation. The runner takes centralized case names and requested region counts as inputs, generates the corresponding distributed datasets, verifies the regional data, and optionally runs the distributed DC OPF and AC OPF validation workflows.

\section{DPLib: Benchmark Test Cases}\label{sec:case_studies}

This section presents a detailed description and verification of several DPLib benchmark systems and their corresponding multi-region topologies. All test cases are partitioned using the spectral clustering procedure of Algorithm~1 based on the unweighted Laplacian matrix. The resulting number of regions, tie-lines, sensitivity of the selected partitions to repeated \(k\)-means calls, and centralized OPF performance are summarized in Table~\ref{tab:kmeans_sensitivity}. Table~\ref{tab:metis_comparison} compares DPLib unweighted partitioning, DPLib weighted partitioning, METIS, the IPA, and KaFFPa on representative benchmark cases. Table~\ref{tab:partition_summary_new} reports the ADMM iteration count, wall clock runtime, and final optimality gap for both distributed DC OPF and AC OPF validation runs and compares DPLib ADMM-based OPF solvers with the PowerModelsADA ADMM-based solver.

To quantify sensitivity, we recorded the tie-line count and feasibility status for each repeated \(k\)-means call in the benchmark cases. Table~\ref{tab:kmeans_sensitivity} reports the number of recorded runs, the number of valid runs, and the minimum, mean, standard deviation, and maximum tie-line counts across valid candidates. In all reported cases, all 2500 recorded candidates were valid, meaning that no repeated run produced an empty region, a generatorless region, or an invalid branch-accounting result. This supports the robustness of the automatic data generation pipeline with respect to feasibility checks imposed by DPLib. For the smaller systems, the tie-line count is identical across all repeated runs. In larger systems, the standard deviation increases, indicating that the final partition is more sensitive to \(k\)-means initialization.

 Table~\ref{tab:kmeans_sensitivity} also reports the centralized DC and AC OPF wall clock run times for the representative benchmark cases. These centralized run times are included to make the computational cost of the reference OPF solve transparent. Together with Table~\ref{tab:partition_summary_new}, which reports the distributed DC and AC OPF iteration counts, wall clock run times, and final optimality gaps, the revised results allow readers to compare the centralized reference solves and the distributed validation workflow. We emphasize that these results are not used to claim that the distributed solvers in DPLib are faster than centralized solvers.

\begin{table*}[!t]
\centering
\captionsetup{font={footnotesize}}
\caption{Centralized OPF performance and sensitivity of the automatic partitioning procedure across repeated recorded \(k\)-means calls}
\label{tab:kmeans_sensitivity}
\scriptsize
\setlength{\tabcolsep}{3.5pt}
\renewcommand{\arraystretch}{1.0}
\resizebox{\textwidth}{!}{%
\begin{tabular}{
l
r
c
r
S[table-format=7.0]
>{\color{black}}S[table-format=3.2]
S[table-format=7.0]
>{\color{black}}S[table-format=3.2]
>{\color{black}}r
>{\color{black}}r
>{\color{black}}r
>{\color{black}}S[table-format=3.2]
>{\color{black}}S[table-format=2.2]
>{\color{black}}r
}
\toprule
& \multicolumn{3}{c}{Partitioning Setup}
& \multicolumn{4}{c}{Centralized OPF}
& \multicolumn{6}{c}{\color{black}Repeated \(k\)-Means Results} \\
\cmidrule(lr){2-4}
\cmidrule(lr){5-8}
\cmidrule(lr){9-14}
\multicolumn{1}{c}{\textbf{Case}} &
\multicolumn{1}{c}{\textbf{Regions}} &
\multicolumn{1}{c}{\textbf{Slack}} &
\multicolumn{1}{c}{\textbf{tie-lines}} &
\multicolumn{1}{c}{\textbf{DC Cost}} &
\multicolumn{1}{c}{\color{black}\textbf{DC Time (s)}} &
\multicolumn{1}{c}{\textbf{AC Cost}} &
\multicolumn{1}{c}{\color{black}\textbf{AC Time (s)}} &
\multicolumn{1}{c}{\color{black}\textbf{Runs}} &
\multicolumn{1}{c}{\color{black}\textbf{Valid}} &
\multicolumn{1}{c}{\color{black}\textbf{Best}} &
\multicolumn{1}{c}{\color{black}\textbf{Mean}} &
\multicolumn{1}{c}{\color{black}\textbf{Std.}} &
\multicolumn{1}{c}{\color{black}\textbf{Worst}} \\
\midrule

\texttt{pglib\_opf\_case200\_tamu} 
& 3  & R2  & 10  
& 34653   & 0.87
& 27558   & 0.42
& 2500 & 2500 & 10  & 10.00  & 0.00  & 10  \\

\texttt{pglib\_opf\_case300\_ieee} 
& 3  & R1  & 12  
& 517310  & 0.65
& 565220  & 0.60
& 2500 & 2500 & 12  & 12.00  & 0.00  & 12  \\

\texttt{pglib\_opf\_case500\_tamu} 
& 5  & R2  & 16  
& 91841   & 0.82
& 72578   & 1.39
& 2500 & 2500 & 16  & 16.00  & 0.06  & 19  \\

\texttt{pglib\_opf\_case1354\_pegase} 
& 8  & R1  & 62  
& 1218028 & 2.90
& 1258844 & 8.54
& 2500 & 2500 & 62  & 64.62  & 2.16  & 75  \\

\texttt{pglib\_opf\_case2869\_pegase} 
& 12 & R6  & 95  
& 2386170 & 8.76
& 2462790 & 50.49
& 2500 & 2500 & 95  & 105.27 & 4.38  & 123 \\

\texttt{pglib\_opf\_case4661\_sdet} 
& 14 & R3  & 284 
& 2217130 & 14.13
& 2255540 & 332.40
& 2500 & 2500 & 284 & 310.18 & 13.47 & 343 \\

\texttt{pglib\_opf\_case9241\_pegase} 
& 18 & R17 & 197 
& 6044174 & 56.09
& 6243090 & 639.44
& 2500 & 2500 & 197 & 229.75 & 10.92 & 289 \\

\bottomrule
\end{tabular}%
}
\end{table*}

\begin{table*}[!t]
\centering
\captionsetup{font={footnotesize}}
\caption{Comparison among DPLib unweighted partitioning, DPLib weighted partitioning, METIS, the IPA, and KaFFPa}
\label{tab:metis_comparison}
\scriptsize
\setlength{\tabcolsep}{1.75pt}
\renewcommand{\arraystretch}{1.12}
\resizebox{\textwidth}{!}{%
\begin{tabular}{
l
S[table-format=2.0]  % regions
S[table-format=3.0]  % Unweighted tie-lines
S[table-format=3.0]  % Weighted tie-lines
S[table-format=3.0]  % METIS tie-lines
S[table-format=3.0]  % IPA tie-lines
S[table-format=3.0]  % KaFFPa tie-lines
S[table-format=3.0]  % Unweighted min bus
S[table-format=3.0]  % Weighted min bus
S[table-format=3.0]  % METIS min bus
S[table-format=3.0]  % IPA min bus
S[table-format=3.0]  % KaFFPa min bus
S[table-format=4.0]  % Unweighted max bus
S[table-format=4.0]  % Weighted max bus
S[table-format=4.0]  % METIS max bus
S[table-format=4.0]  % IPA max bus
S[table-format=4.0]  % KaFFPa max bus
S[table-format=2.2]  % Unweighted balance
S[table-format=2.2]  % Weighted balance
S[table-format=2.2]  % METIS balance
S[table-format=2.2]  % IPA balance
S[table-format=2.2]  % KaFFPa balance
}
\toprule
&
& \multicolumn{5}{c}{tie-lines}
& \multicolumn{5}{c}{Min Bus}
& \multicolumn{5}{c}{Max Bus}
& \multicolumn{5}{c}{Bus-Balance Ratio} \\
\cmidrule(lr){3-7}
\cmidrule(lr){8-12}
\cmidrule(lr){13-17}
\cmidrule(lr){18-22}
\multicolumn{1}{c}{\textbf{Case}} &
\multicolumn{1}{c}{\textbf{Regions}} &
\multicolumn{1}{c}{\textbf{Unwei.}} &
\multicolumn{1}{c}{\textbf{Wei.}} &
\multicolumn{1}{c}{\textbf{METIS}} &
\multicolumn{1}{c}{\textbf{IPA-insp.}} &
\multicolumn{1}{c}{\textbf{KaFFPa}} &
\multicolumn{1}{c}{\textbf{Unwei.}} &
\multicolumn{1}{c}{\textbf{Wei.}} &
\multicolumn{1}{c}{\textbf{METIS}} &
\multicolumn{1}{c}{\textbf{IPA-insp.}} &
\multicolumn{1}{c}{\textbf{KaFFPa}} &
\multicolumn{1}{c}{\textbf{Unwei.}} &
\multicolumn{1}{c}{\textbf{Wei.}} &
\multicolumn{1}{c}{\textbf{METIS}} &
\multicolumn{1}{c}{\textbf{IPA-insp.}} &
\multicolumn{1}{c}{\textbf{KaFFPa}} &
\multicolumn{1}{c}{\textbf{Unwei.}} &
\multicolumn{1}{c}{\textbf{Wei.}} &
\multicolumn{1}{c}{\textbf{METIS}} &
\multicolumn{1}{c}{\textbf{IPA-insp.}} &
\multicolumn{1}{c}{\textbf{KaFFPa}} \\
\midrule

\texttt{ 200\_tamu}
& 3
& 10 & 12 & 11 & 8 & 11
& 49 & 21 & 66 & 12 & 66
& 87 & 109 & 67 & 134 & 67
& 1.78 & 5.19 & 1.02 & 11.17 & 1.02 \\

\texttt{ 500\_tamu}
& 5
& 16 & 17 & 22 & 15 & 20
& 47 & 43 & 97 & 28 & 99
& 205 & 211 & 103 & 191 & 102
& 4.36 & 4.91 & 1.06 & 6.82 & 1.03 \\

\texttt{ 1354\_pegase}
& 8
& 62 & 62 & 89 & 65 & 70
& 68 & 68 & 164 & 30 & 159
& 289 & 289 & 174 & 231 & 174
& 4.25 & 4.25 & 1.06 & 7.70 & 1.09 \\

\texttt{ 4661\_sdet}
& 14
& 284 & 277 & 286 & 309 & 258
& 79 & 75 & 328 & 23 & 309
& 613 & 590 & 339 & 756 & 341
& 7.76 & 7.87 & 1.03 & 32.87 & 1.10 \\

\texttt{ 9241\_pegase}
& 18
& 197 & 197 & 289 & 480 & 235
& 145 & 145 & 498 & 56 & 482
& 851 & 851 & 527 & 2008 & 528
& 5.87 & 5.87 & 1.06 & 35.86 & 1.10 \\

\bottomrule
\end{tabular}%
}
\end{table*}

\begin{table*}[!t]
\centering
\captionsetup{font={footnotesize}}
\caption{Distributed OPF validation metrics and software-level comparison with PowerModelsADA}
\label{tab:partition_summary_new}
\scriptsize
\setlength{\tabcolsep}{1.8pt}
\renewcommand{\arraystretch}{1.12}
\resizebox{\textwidth}{!}{%
\begin{tabular}{
l
c
c
c
c
c
c
c
c
c
c
c
c
c
c
c
}
\toprule
& \multicolumn{3}{c}{Penalty Param.} 
& \multicolumn{6}{c}{\textcolor{black}{Distributed OPF with DPLib}}
& \multicolumn{6}{c}{\textcolor{black}{Distributed OPF with PowerModelsADA}} \\
\cmidrule(lr){2-4}
\cmidrule(lr){5-10}
\cmidrule(lr){11-16}
\multicolumn{1}{c}{\textbf{Case}} &
\multicolumn{1}{c}{\boldmath$\rho$} &
\multicolumn{1}{c}{\boldmath$\rho_{\mathrm{DC}}^{\max}$} &
\multicolumn{1}{c}{\boldmath$\rho_{\mathrm{AC}}^{\max}$} &
\multicolumn{1}{c}{\textcolor{black}{\textbf{DC Iter.}}} &
\multicolumn{1}{c}{\textcolor{black}{\textbf{DC Time (s)}}} &
\multicolumn{1}{c}{\textcolor{black}{\textbf{DC Gap (\%)}}} &
\multicolumn{1}{c}{\textcolor{black}{\textbf{AC Iter.}}} &
\multicolumn{1}{c}{\textcolor{black}{\textbf{AC Time (s)}}} &
\multicolumn{1}{c}{\textcolor{black}{\textbf{AC Gap (\%)}}} &
\multicolumn{1}{c}{\textcolor{black}{\textbf{DC Iter.}}} &
\multicolumn{1}{c}{\textcolor{black}{\textbf{DC Time (s)}}} &
\multicolumn{1}{c}{\textcolor{black}{\textbf{DC Gap (\%)}}} &
\multicolumn{1}{c}{\textcolor{black}{\textbf{AC Iter.}}} &
\multicolumn{1}{c}{\textcolor{black}{\textbf{AC Time (s)}}} &
\multicolumn{1}{c}{\textcolor{black}{\textbf{AC Gap (\%)}}} \\
\midrule

\texttt{200\_tamu}    
& 1 & 1e8 & 1e8 
& \textcolor{black}{473} & \textcolor{black}{226} & \textcolor{black}{2e-2}
& \textcolor{black}{461} & \textcolor{black}{214} & \textcolor{black}{4e-2}
& \textcolor{black}{418} & \textcolor{black}{22.14} & \textcolor{black}{813.03}
& \textcolor{black}{707} & \textcolor{black}{109.45} & \textcolor{black}{843.15} \\

\texttt{300\_ieee}    
& 1 & 1e8 & 1e8 
& \textcolor{black}{523} & \textcolor{black}{295} & \textcolor{black}{8e-3}
& \textcolor{black}{444} & \textcolor{black}{286} & \textcolor{black}{1e-2}
& \textcolor{black}{71} & \textcolor{black}{6.07} & \textcolor{black}{1567.50}
& \textcolor{black}{130} & \textcolor{black}{39.43} & \textcolor{black}{1094.35} \\

\texttt{500\_tamu}    
& 1 & 1e8 & 1e8 
& \textcolor{black}{418} & \textcolor{black}{390} & \textcolor{black}{4e-3}
& \textcolor{black}{418} & \textcolor{black}{414} & \textcolor{black}{1e-1}
& \textcolor{black}{1217} & \textcolor{black}{75.78} & \textcolor{black}{2046.11}
& \textcolor{black}{1553} & \textcolor{black}{557.31} & \textcolor{black}{1815.38} \\

\texttt{1354\_pegase} 
& 1 & 1e8 & 1e8 
& \textcolor{black}{466} & \textcolor{black}{1201} & \textcolor{black}{6e-3}
& \textcolor{black}{427} & \textcolor{black}{2402} & \textcolor{black}{2e-2}
& \textcolor{black}{4295} & \textcolor{black}{814.08} & \textcolor{black}{1254.57}
& \textcolor{black}{$10000^{\dagger}$} & \textcolor{black}{23591.91} & \textcolor{black}{1206.13} \\

\texttt{2869\_pegase} 
& 1 & 1e5 & 1e8 
& \textcolor{black}{1380} & \textcolor{black}{3200} & \textcolor{black}{5e-2}
& \textcolor{black}{1155} & \textcolor{black}{5200} & \textcolor{black}{9e-1}
& \textcolor{black}{$10000^{\dagger}$} & \textcolor{black}{7015.92} & \textcolor{black}{12.02}
& \textcolor{black}{$10000^{\dagger}$} & \textcolor{black}{40787.80} & \textcolor{black}{890.08} \\

\texttt{4661\_sdet}   
& 1 & 1e6 & 1e6 
& \textcolor{black}{5713} & \textcolor{black}{12800} & \textcolor{black}{5e-1}
& \textcolor{black}{7352} & \textcolor{black}{28400} & \textcolor{black}{3e-1}
& \textcolor{black}{$10000^{\dagger}$} & \textcolor{black}{23065.23} & \textcolor{black}{4.48}
& \textcolor{black}{$10000^{\dagger}$} & \textcolor{black}{45444.85} & \textcolor{black}{20.85} \\

\texttt{9241\_pegase} 
& 1 & 1e6 & 1e5 
& \textcolor{black}{1702} & \textcolor{black}{7600} & \textcolor{black}{7e-1}
& \textcolor{black}{793} & \textcolor{black}{9400} & \textcolor{black}{8e-1}
& \textcolor{black}{$10000^{\dagger}$} & \textcolor{black}{43191.54} & \textcolor{black}{92.35}
& \textcolor{black}{$10000^{\dagger}$} & \textcolor{black}{162257.07} & \textcolor{black}{0.93} \\

\bottomrule
\end{tabular}%
}
\vspace{1mm}
\begin{flushleft}
\scriptsize
\(\dagger\) indicates that the PowerModelsADA run reached the maximum ADMM iteration limit. 
\end{flushleft}
\end{table*}

\begin{table*}[!t]
\centering
\captionsetup{font={footnotesize}}
\caption{Impact of partitioning method on distributed AC OPF convergence performance}
\label{tab:partition_method_comparison}
\scriptsize
\setlength{\tabcolsep}{2.8pt}
\renewcommand{\arraystretch}{1.0}
\resizebox{\textwidth}{!}{%
\begin{tabular}{
l
r
>{\color{black}}r
>{\color{black}}S[table-format=4.0]
>{\color{black}}S[table-format=4.2]
>{\color{black}}c
>{\color{black}}r
>{\color{black}}S[table-format=4.0]
>{\color{black}}S[table-format=4.2]
>{\color{black}}c
>{\color{black}}r
>{\color{black}}S[table-format=4.0]
>{\color{black}}S[table-format=4.2]
>{\color{black}}c
}
\toprule
& \multicolumn{1}{c}{  }
& \multicolumn{4}{c}{\color{black}\textbf{DPLib}} 
& \multicolumn{4}{c}{\color{black}\textbf{METIS}} 
& \multicolumn{4}{c}{\color{black}\textbf{KaFFPa}} \\
\cmidrule(lr){2-2}
\cmidrule(lr){3-6}
\cmidrule(lr){7-10}
\cmidrule(lr){11-14}
\multicolumn{1}{c}{\textbf{Case}} &
\multicolumn{1}{c}{\textbf{Regions}} &
\multicolumn{1}{c}{\color{black}\textbf{Tie-lines}} &
\multicolumn{1}{c}{\color{black}\textbf{Iter.}} &
\multicolumn{1}{c}{\color{black}\textbf{Time (s)}} &
\multicolumn{1}{c}{\color{black}\textbf{Gap}} &
\multicolumn{1}{c}{\color{black}\textbf{Tie-lines}} &
\multicolumn{1}{c}{\color{black}\textbf{Iter.}} &
\multicolumn{1}{c}{\color{black}\textbf{Time (s)}} &
\multicolumn{1}{c}{\color{black}\textbf{Gap}} &
\multicolumn{1}{c}{\color{black}\textbf{Tie-lines}} &
\multicolumn{1}{c}{\color{black}\textbf{Iter.}} &
\multicolumn{1}{c}{\color{black}\textbf{Time (s)}} &
\multicolumn{1}{c}{\color{black}\textbf{Gap}} \\
\midrule

\texttt{pglib\_opf\_case30\_ieee} 
& 2 
& 4  & 126  & 30   & 4e-2 
& 6  & 160  & 37   & 1e-1   
& 5  & 147  & 34   & 7e-2 \\

\texttt{pglib\_opf\_case118\_ieee} 
& 2 
& 5  & 136  & 45   & 4e-3  
& 7  & 230  & 78   & 7e-3   
& 7  & 156  & 53   & 2e-3  \\

\texttt{pglib\_opf\_case118\_ieee} 
& 3 
& 10 & 255  & 116  & 3e-0  
& 11 & 172  & 77   & 6e-2   
& 9  & 150  & 66   & 2e-3  \\

\texttt{pglib\_opf\_case118\_ieee} 
& 4 
& 14 & 108  & 60   & 1e-1  
& 18 & 208  & 119  & 7e-1   
& 16 & 166  & 93   & 1e-1  \\

\texttt{pglib\_opf\_case200\_tamu} 
& 2 
& 8  & 285  & 93   & 1e-1  
& 7  & 273  & 87   & 6e-2   
& 7  & 299  & 96   & 7e-2 \\

\texttt{pglib\_opf\_case200\_tamu} 
& 3 
& 10 & 220  & 99   & 9e-2 
& 11 & 222  & 93   & 8e-2   
& 11 & 265  & 112  & 9e-2  \\

\texttt{pglib\_opf\_case200\_tamu} 
& 4 
& 13 & 224  & 121  & 1e-1  
& 15 & 246  & 137  & 1e-1   
& 14 & 231  & 123  & 1e-1  \\

\texttt{pglib\_opf\_case300\_ieee} 
& 3 
& 12 & 282  & 182  & 8e-3  
& 9  & 183  & 118  & 3e-3   
& 9  & 204  & 129  & 1e-2  \\

\texttt{pglib\_opf\_case300\_ieee} 
& 4 
& 11 & 1955 & 1906 & 6e-0  
& 17 & 1156 & 982  & 2e+1  
& 14 & 5000 & 3358 & 7e-0 \\

\texttt{pglib\_opf\_case500\_tamu} 
& 4 
& 14 & 286  & 239  & 5e-2  
& 19 & 308  & 253  & 1e-1   
& 16 & 239  & 189  & 3e-2  \\

\texttt{pglib\_opf\_case500\_tamu} 
& 5 
& 16 & 213  & 194  & 7e-2  
& 22 & 261  & 234  & 6e-2   
& 20 & 270  & 245  & 1e-1  \\

\texttt{pglib\_opf\_1354\_pegase} 
& 6 
& 51 & 470  & 1217  & 2e-1  
& 86 & 288  & 857  & 6e-2   
& 63 & 298  & 1116  & 2e-1  \\

\texttt{pglib\_opf\_1354\_pegase} 
& 7 
& 58 & 275  & 745  & 2e-2  
& 87 & 235  & 644  & 7e-1   
& 62 & 366  & 1007  & 3e-1  \\

\texttt{pglib\_opf\_1354\_pegase} 
& 8 
& 62 & 251  & 693  & 2e-1  
& 89 & 318  & 864  & 2e-0   
& 70 & 159  & 409  & 8e-3  \\

\bottomrule
\end{tabular}%
}
\end{table*}

 To further position the proposed automatic partitioning procedure, we compare five alternatives on representative benchmark cases: the default DPLib unweighted partitioning, the DPLib weighted partitioning, METIS, an IPA-inspired coupling-based baseline, and KaFFPa. The IPA-inspired baseline follows the main workflow of the Intelligent Partition Algorithm in~\cite{guo2015intelligent}, but it is not an exact reproduction because the original implementation is not publicly available. KaFFPa is included as an additional high-quality graph partitioning reference. Table~\ref{tab:metis_comparison} reports the number of inter-regional tie-lines and the regional bus count balance for these alternatives. Graph partitioning is measured as \(\max_r |\mathcal{V}_r|/\min_r |\mathcal{V}_r|\), where values closer to 1 indicate more equal region sizes. 

The comparison highlights the different roles and behaviors of the five alternatives. METIS produces highly balanced graph partitions, with bus balance ratios close to 1 across all representative cases, consistent with its balanced graph partitioning objective. KaFFPa also produces highly balanced partitions, with bus balance ratios close to one across all cases. This behavior is consistent with the role of KaFFPa as a multilevel graph partitioner designed to generate balanced partitions with a small number of cut edges, and it also agrees with the motivation for using KaFFPa in~\cite{kyesswa2020impact}. Compared with METIS, KaFFPa provides a similar bus count balance but often reduces the number of tie-lines. Thus, KaFFPa provides a strong balanced graph partitioning baseline that is especially competitive on the larger cases.

The DPLib unweighted mode produces fewer size-balanced regions than METIS and KaFFPa, but it gives fewer inter-regional tie-lines than METIS in all reported cases and fewer tie-lines than KaFFPa in several cases. This supports its use as the default topology-based benchmark generation mode, where the goal is to obtain coherent regional datasets with fewer boundary connections rather than nearly equal bus counts. The DPLib weighted mode shows the effect of including electrical line information in the adjacency matrix. For smaller cases, the weighted mode does not improve the tie-line count relative to the unweighted default, and in some cases it produces less balanced regions. However, for larger systems such as \texttt{case4661\_sdet} and \texttt{case9241\_pegase}, the weighted mode slightly reduces the number of tie-lines compared with the unweighted mode. In \texttt{case9241\_pegase}, the weighted DPLib mode gives the smallest tie-line count among the reported alternatives, although its graph partitioning is weaker than METIS and KaFFPa. This shows that the weighted formulation can be useful when users want electrical parameters to influence the partition.

The IPA-inspired baseline provides a different, coupling-aware reference. In smaller cases, it produces fewer tie-lines than the DPLib, METIS, and KaFFPa alternatives. For example, in \texttt{case200\_tamu} and \texttt{case500\_tamu}, it gives the fewest tie-lines among the reported alternatives. However, it also produces more uneven region sizes, as reflected by larger bus balance ratios. This behavior is expected because the IPA-inspired baseline is driven by coupling-aware clustering rather than bus count balance or benchmark data construction. For the larger systems, the IPA-inspired baseline produces significantly fewer balanced regions and more tie-lines than DPLib, METIS, and KaFFPa, especially for \texttt{case4661\_sdet} and \texttt{case9241\_pegase}. These results show that coupling-aware partitioning can behave differently from topology-based benchmark partitioning and balanced graph partitioning, and that the most suitable partition depends on the intended use.

More importantly, the DPLib variants directly produce the distributed data required for benchmarking, including regional MATPOWER-compatible cases, local bus numbering, generator and cost mappings, explicit tie-line records, and topology outputs. By contrast, METIS, KaFFPa, and the IPA-inspired baseline provide only bus-to-region labels and require additional power-system-specific preprocessing before they can be used as distributed benchmark cases. Thus, DPLib is not proposed as a universal replacement for graph partitioners or OPF-oriented partitioning methods; it addresses the distinct task of converting centralized MATPOWER-compatible systems into complete, inspectable, and reusable multi-region datasets. When users prefer an external partition, such as a METIS, KaFFPa, IPA-inspired, market-zone, control-area, or institutionally defined partition, DPLib can import the corresponding bus-to-region assignment and convert it into the same standardized data format.

To make the library easy to use and to avoid case by case tuning, the 
ADMM penalty parameters are initialized to $\rho^{0}=1$ for all 
benchmarks in both the distributed DC OPF and AC OPF solvers, after 
which $\rho$ is adjusted automatically via the adaptive residual 
balancing rule where \(  \;
  \tau_{\mathrm{incr}}=2,
  \tau_{\mathrm{decr}}=0.5\). The values 
$\rho_{\mathrm{DC}}^{\max}$ and $\rho_{\mathrm{AC}}^{\max}$ in 
Table~\ref{tab:partition_summary_new} are upper caps on the adaptive 
update, not fixed penalty values; the penalty starts at $\rho^{0}=1$ 
and reaches the cap only if the adaptive rule pushes it there. In the distributed DC OPF solver, the adaptive update parameters are set internally to 
\begin{align}
     \mu=\max\!\left(1,\ \frac{15}{(n_{\mathrm{reg}}^{2})\cdot 0.2}\right),  
\end{align}

where $n_{\mathrm{reg}}$ denotes the number of regions. For the distributed AC OPF solver, the same heuristic is applied separately to the angle and voltage magnitude channels using
\begin{align}
     &\mu_{\theta}=\max\!\left(1,\ \frac{15}{(n_{\mathrm{reg}}^{2})\cdot 0.2}\right), 
  \;
  \mu_{v}=\max\!\left(1,\ \frac{10}{(n_{\mathrm{reg}}^{2})\cdot 0.2}\right).
\end{align}

These choices scale the residual balance thresholds with the partition granularity while keeping the increase/decrease factors fixed and interpretable, allowing the adaptive-$\rho$ mechanism to operate robustly across different network sizes and region counts with minimal user intervention.

{\color{black}
Table~\ref{tab:partition_summary_new} reports the distributed OPF validation metrics for the DPLib validation solvers and a software-level comparison with PowerModelsADA using the exported DPLib partition maps. This comparison is not intended as a strict algorithmic benchmark because the two tools use different data structures, modeling layers, solver implementations, and ADMM penalty-update mechanisms. In DPLib, the penalty parameter \(\rho\) is updated adaptively up to the reported maximum value. In PowerModelsADA, the ADMM penalty parameter \(\alpha\) is fixed during each run. For the comparison reported in Table~\ref{tab:partition_summary_new}, \(\alpha\) was set equal to the corresponding DPLib maximum penalty value, i.e., \(\rho_{\mathrm{DC}}^{\max}\) for DC OPF and \(\rho_{\mathrm{AC}}^{\max}\) for AC OPF.

The results show that PowerModelsADA can run on the exported DPLib partition maps, which confirms that the DPLib partition outputs can be used by an external distributed-OPF framework. However, the results also show that using large fixed penalty values in PowerModelsADA can lead to a strong mismatch-reduction behavior without necessarily producing small OPF objective gaps. For example, in the smaller and medium-sized cases, several PowerModelsADA runs satisfy the residual-based convergence criterion, but the final objective gaps remain very large. In \texttt{200\_tamu}, the PowerModelsADA gaps are \(813.03\%\) for DC OPF and \(843.15\%\) for AC OPF, while the corresponding DPLib gaps are \(2\times10^{-2}\%\) and \(4\times10^{-2}\%\). Similar behavior is observed in \texttt{300\_ieee}, \texttt{500\_tamu}, and \texttt{1354\_pegase}, where PowerModelsADA reduces the residual but reports objective gaps above \(1000\%\) for several runs.

For larger systems, PowerModelsADA more often reaches the maximum iteration limit. For \texttt{2869\_pegase}, \texttt{4661\_sdet}, and \texttt{9241\_pegase}, both DC or AC runs reach the \(10000\)-iteration limit in several cases. Some of these runs produce moderate or small final gaps, such as the AC case of \texttt{9241\_pegase}, but they do not satisfy the stopping criterion within the tested iteration limit. These results illustrate the sensitivity of fixed-penalty ADMM implementations to the penalty parameter and network partition. In contrast, the DPLib validation solvers produce consistently small final gaps on all reported cases using the adaptive-\(\rho\) update.

Table~\ref{tab:partition_method_comparison} shows that the partitioning method affects distributed AC OPF behavior, but no single method is uniformly best across all metrics. DPLib produces fewer inter-regional tie-lines in many cases, reducing the number of boundary couplings in the distributed formulation, and it also yields competitive iteration counts and solve times in several cases. However, fewer tie-lines do not always lead to the lowest optimality gap or fastest convergence. METIS and KaFFPa sometimes achieve lower iteration counts, shorter solve times, or smaller final gaps. For example, KaFFPa performs best for \texttt{case118\_ieee\_k3}, while METIS performs best for \texttt{case300\_ieee\_k3}. The difficult \texttt{case300\_ieee\_k4} case also shows that partitioning can strongly affect numerical behavior, but that none of the tested methods is uniformly superior. These results support the intended role of DPLib as a benchmark-data and validation resource: it enables users to generate distributed OPF cases and directly evaluate how different partitioning choices affect tie-line count, ADMM iteration count, solve time, and final optimality gap.}

\subsection*{Case pglib\_opf\_case200\_tamu}

\begin{figure}[t!]
\captionsetup{font={footnotesize}}
\centering
\includegraphics[width=1\columnwidth]{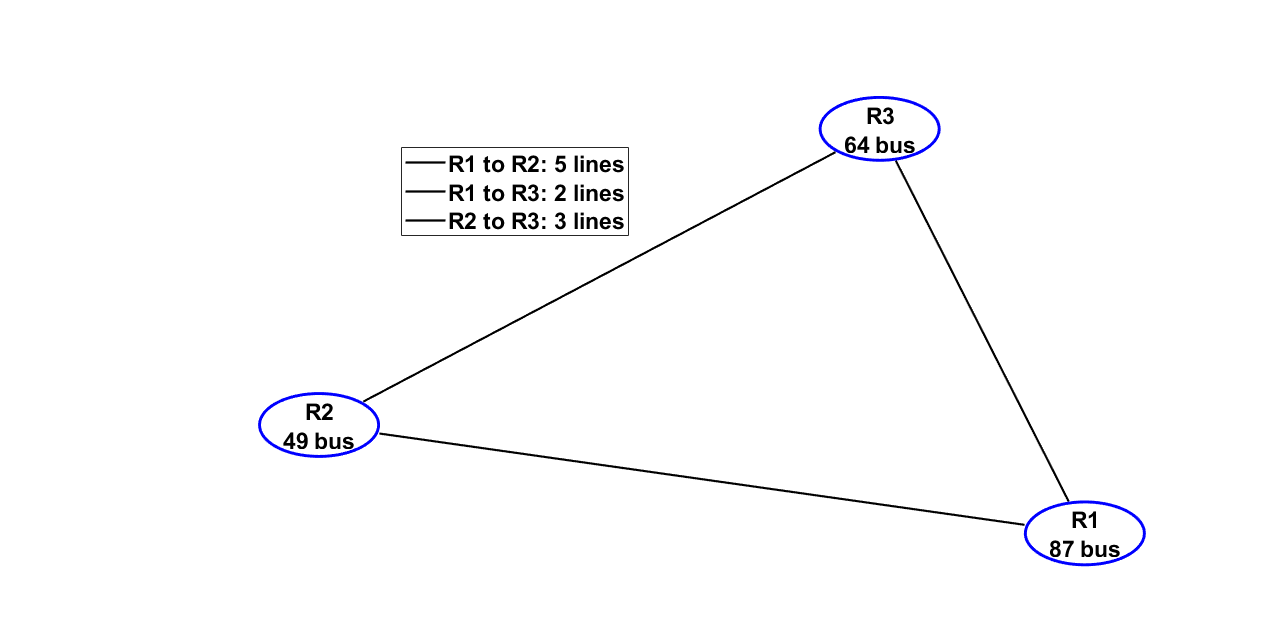}
\caption{Three-region topology for case200\_tamu}
\label{fig:topo_200}
\end{figure}

The system \texttt{pglib\_opf\_case200\_tamu} is a 200-bus synthetic model representing a 230/115 kV network in Illinois. It was generated as part of the ARPA-E GRID DATA project and is designed to capture realistic transmission characteristics and operational patterns. The system is partitioned into three regions connected by ten tie-lines. The resulting multi-region topology is shown in Fig.~\ref{fig:topo_200}.

\begin{figure}[t!]
\captionsetup{font={footnotesize}}
\centering
\includegraphics[width=0.45\linewidth]{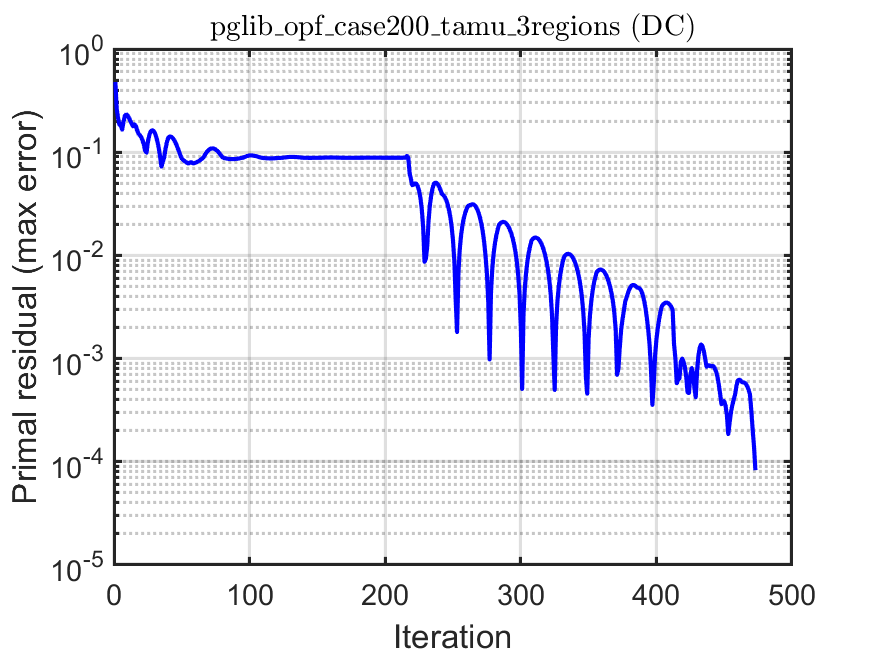}
\includegraphics[width=0.45\linewidth]{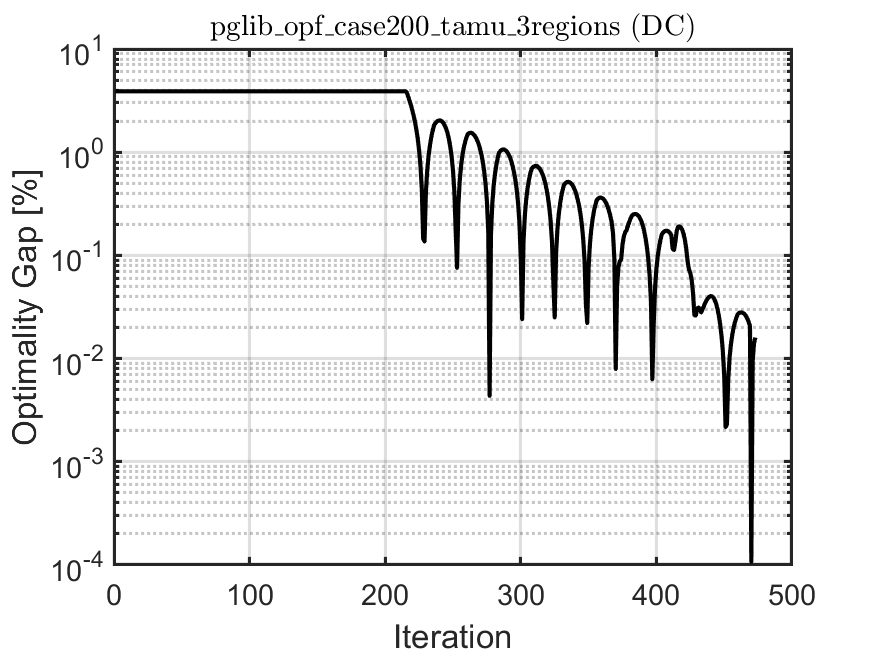}
\caption{primal residual (left) and optimality gap (right)}
\label{fig:case200_dc}
\end{figure}

\begin{figure}[t!]
\captionsetup{font={footnotesize}}
\centering
\includegraphics[width=0.45\linewidth]{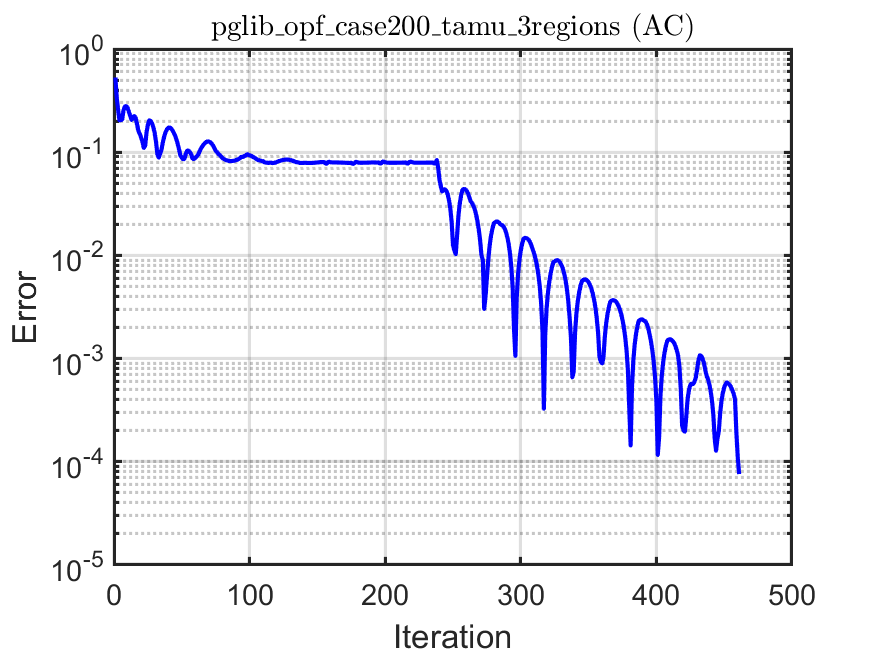}
\includegraphics[width=0.45\linewidth]{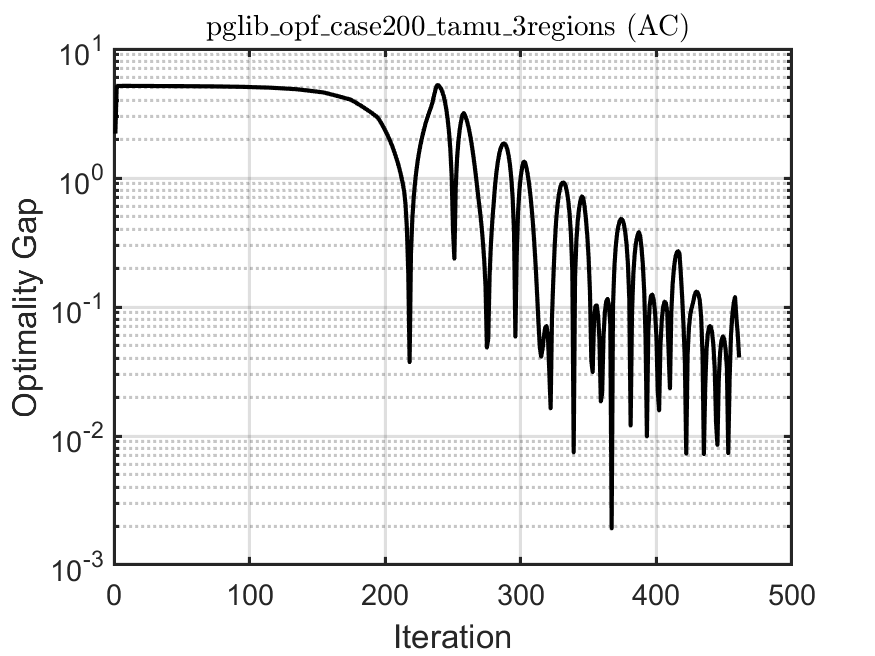}
\caption{primal residual (left) and optimality gap (right)}
\label{fig:case200_ac}
\end{figure}

Figures~\ref{fig:case200_dc} and~\ref{fig:case200_ac} report the DC and AC primal residuals and optimality gaps for \texttt{pglib\_opf\_case200\_tamu}.

\subsection*{Case pglib\_opf\_case300\_ieee}

\begin{figure}[t!]
\captionsetup{font={footnotesize}}
\centering
\includegraphics[width=1\columnwidth]{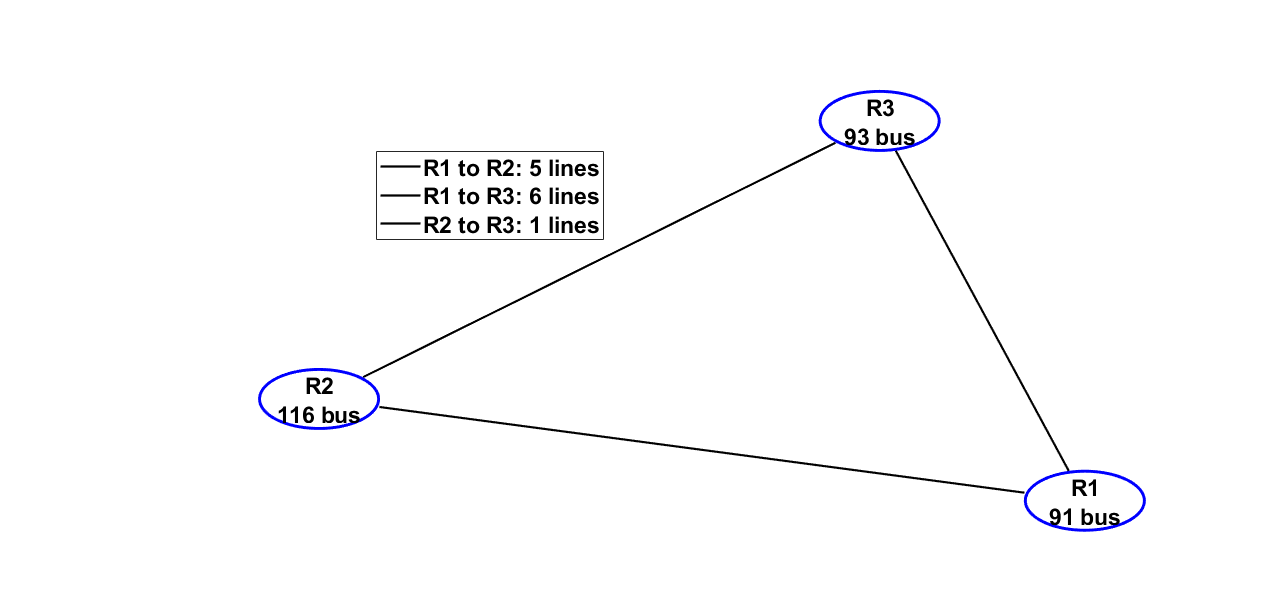}
\caption{Three-region topology for case300\_ieee}
\label{fig:topo_300}
\end{figure}

The system \texttt{pglib\_opf\_case300\_ieee} corresponds to the well-known IEEE 300-bus test case. Originally formulated in the IEEE Common Data Format and subsequently converted to MATPOWER, it has long served as a standard benchmark for power system optimization studies. The system is divided into three regions, with interconnections distributed across 12 tie-lines. The extracted topology is shown in Fig.~\ref{fig:topo_300}.

\begin{figure}[t!]
\captionsetup{font={footnotesize}}
\centering
\includegraphics[width=0.45\linewidth]{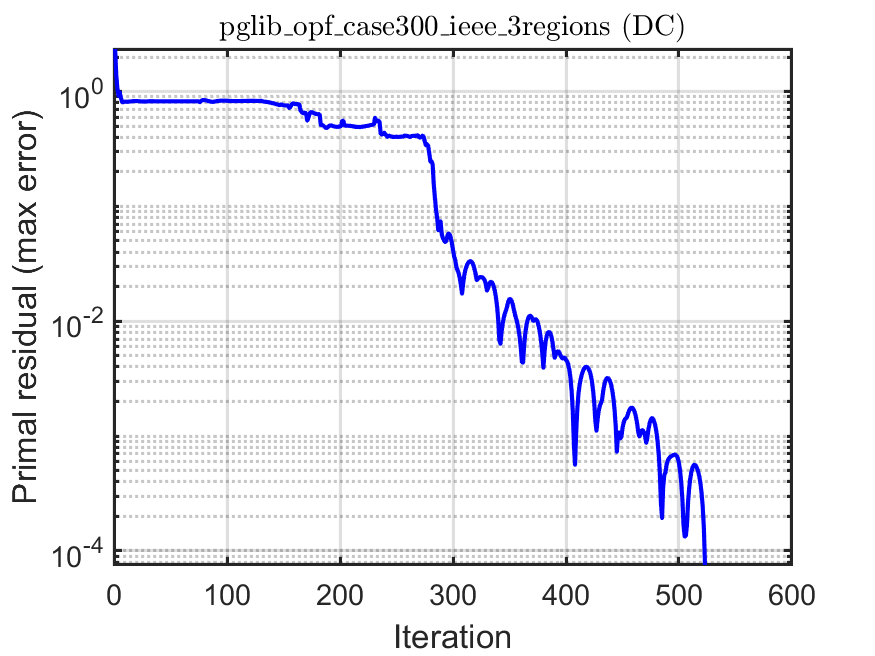}
\includegraphics[width=0.45\linewidth]{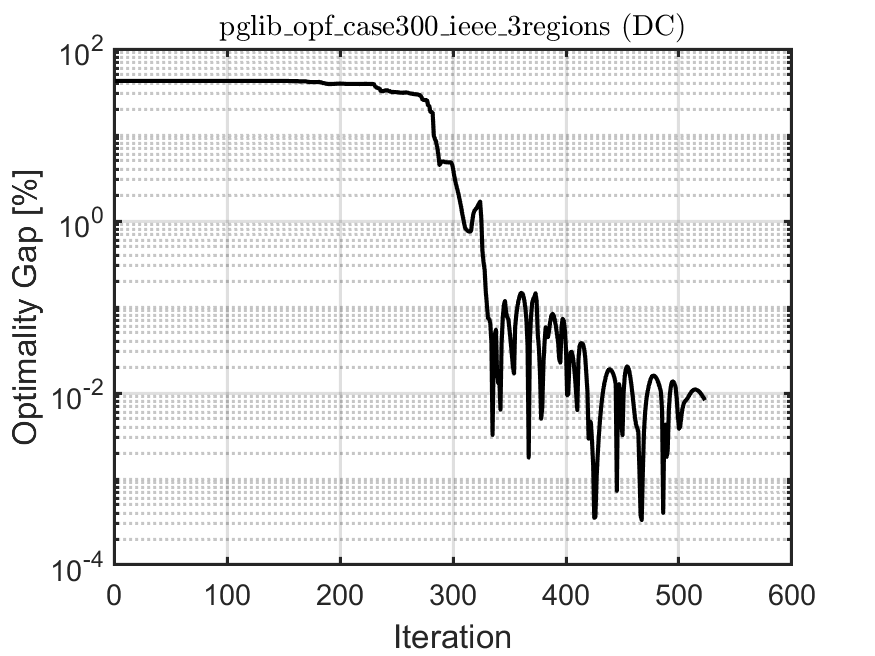}
\caption{primal residual (left) and optimality gap (right)}
\label{fig:case300_dc}
\end{figure}

\begin{figure}[t!]
\captionsetup{font={footnotesize}}
\centering
\includegraphics[width=0.45\linewidth]{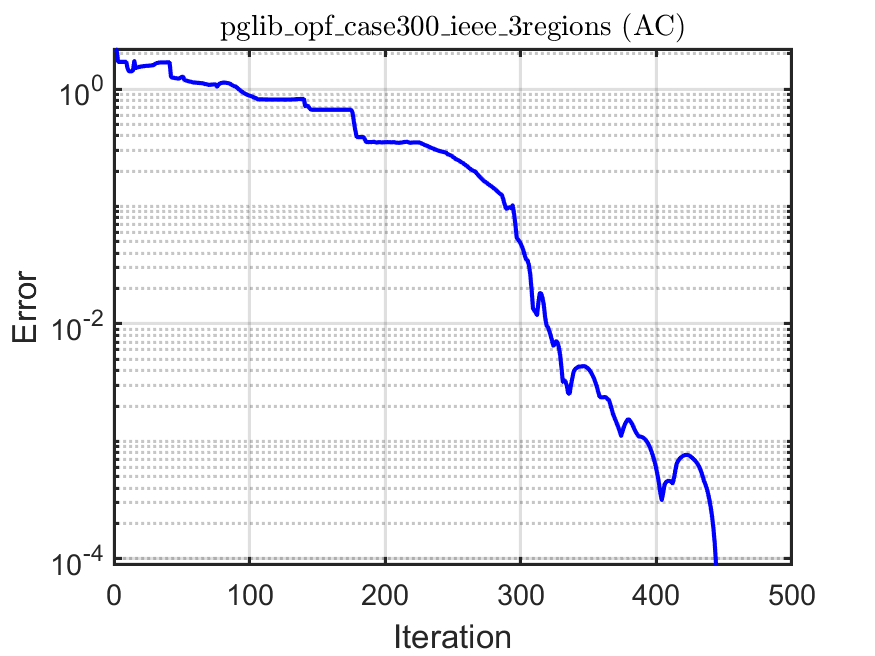}
\includegraphics[width=0.45\linewidth]{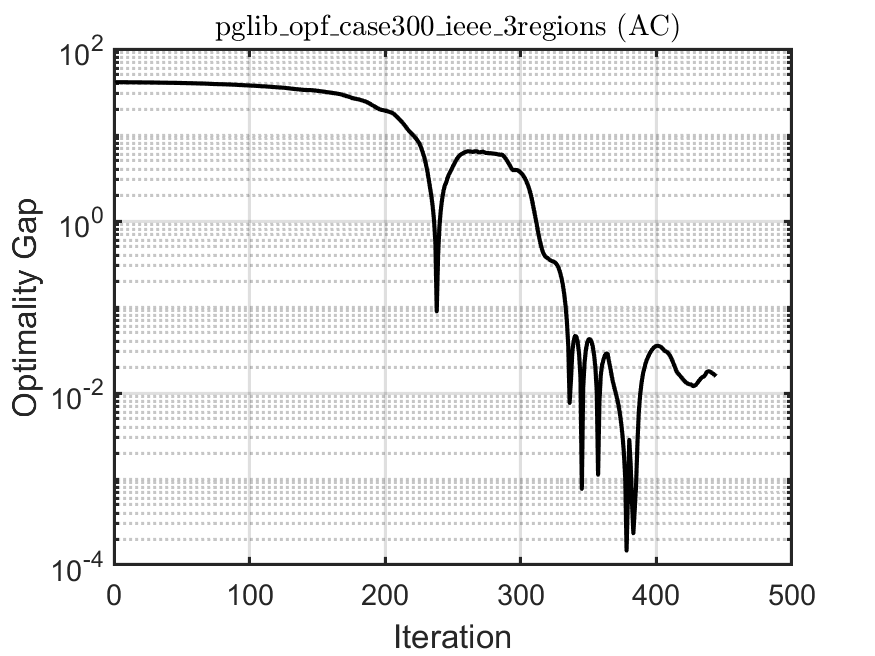}
\caption{primal residual (left) and optimality gap (right)}
\label{fig:case300_ac}
\end{figure}

Figures~\ref{fig:case300_dc} and~\ref{fig:case300_ac} show the DC and AC verification results for \texttt{pglib\_opf\_case300\_ieee}. 

\subsection*{Case pglib\_opf\_case500\_tamu}

\begin{figure}[t!]
\captionsetup{font={footnotesize}}
\centering
\includegraphics[width=1\columnwidth]{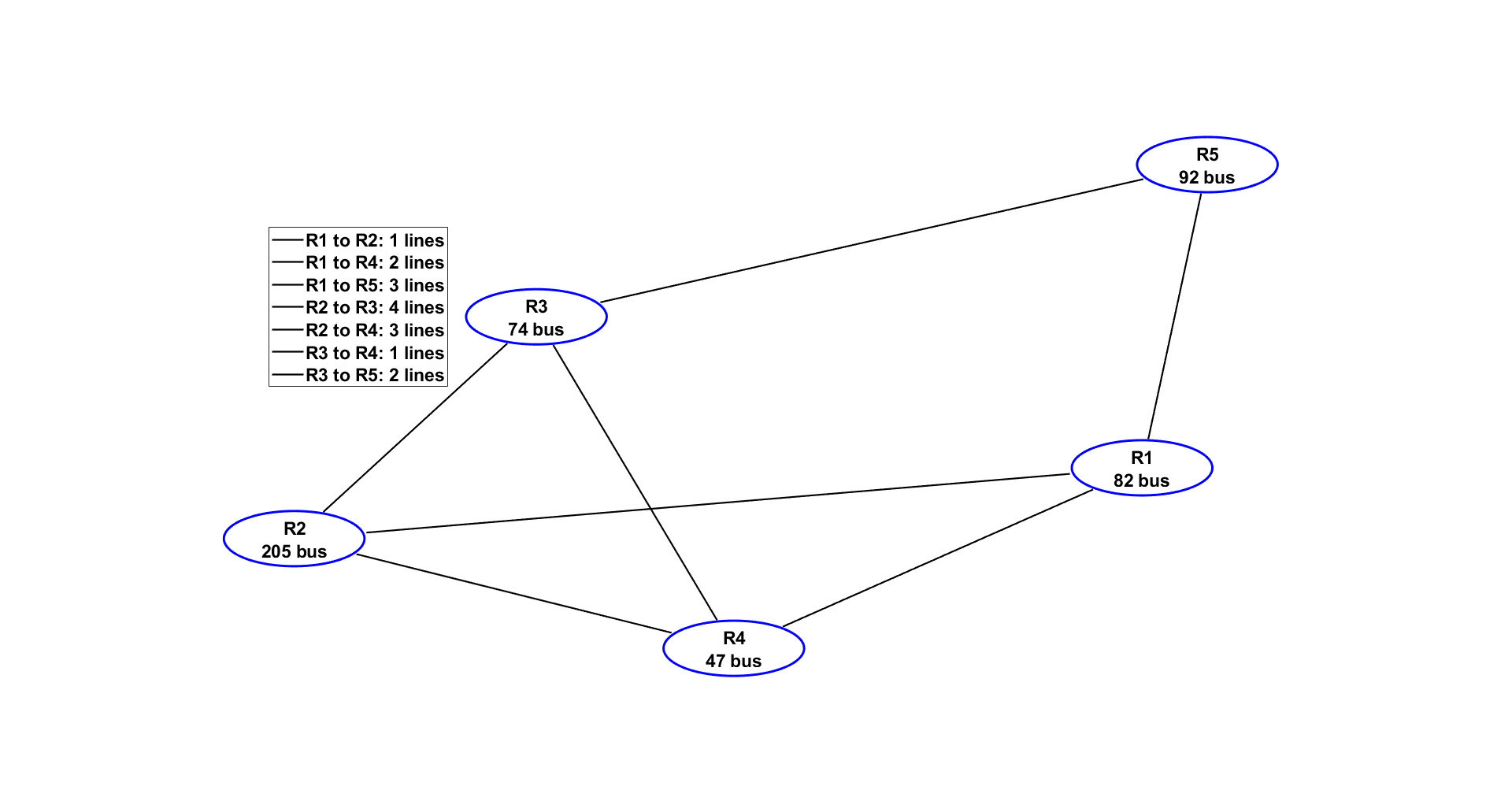}
\caption{Five-region topology for case500\_tamu}
\label{fig:topo_500}
\end{figure}

The system \texttt{pglib\_opf\_case500\_tamu} is a 500-bus synthetic model of the South Carolina transmission grid, also produced under the ARPA-E GRID DATA initiative. It contains detailed line, transformer, and generator data reflecting real-world system behavior. The network is partitioned into five regions connected by sixteen tie-lines. The corresponding topology appears in Fig.~\ref{fig:topo_500}.

\begin{figure}[t!]
\captionsetup{font={footnotesize}}
\centering
\includegraphics[width=0.45\linewidth]{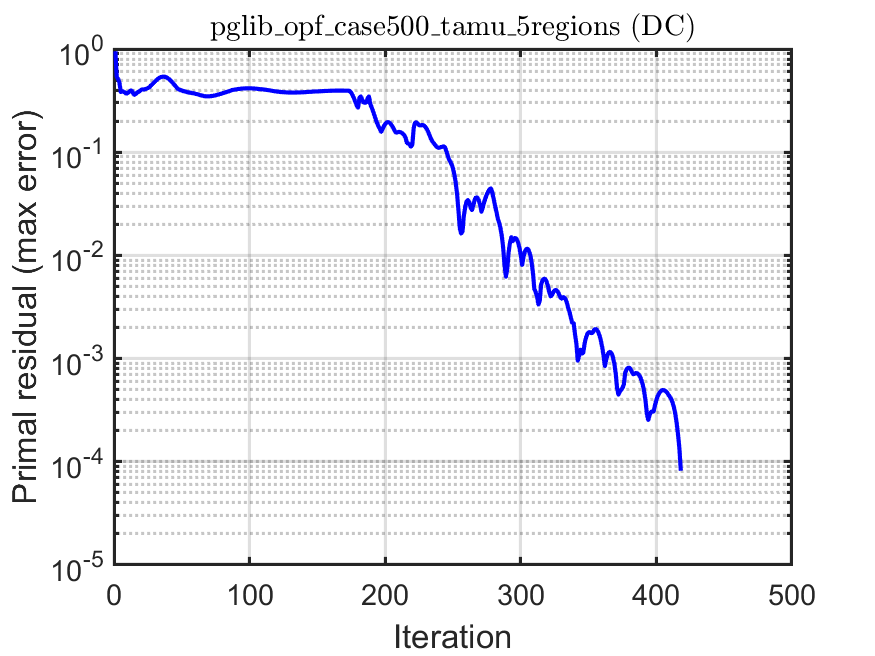}
\includegraphics[width=0.45\linewidth]{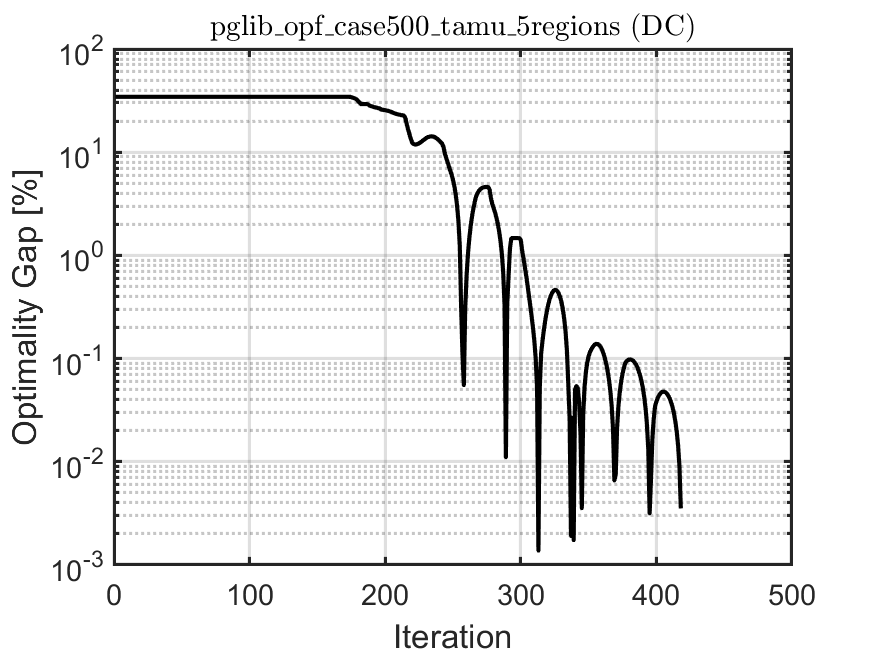}
\caption{primal residual (left) and optimality gap (right)}
\label{fig:case500_dc}
\end{figure}

\begin{figure}[t!]
\captionsetup{font={footnotesize}}
\centering
\includegraphics[width=0.45\linewidth]{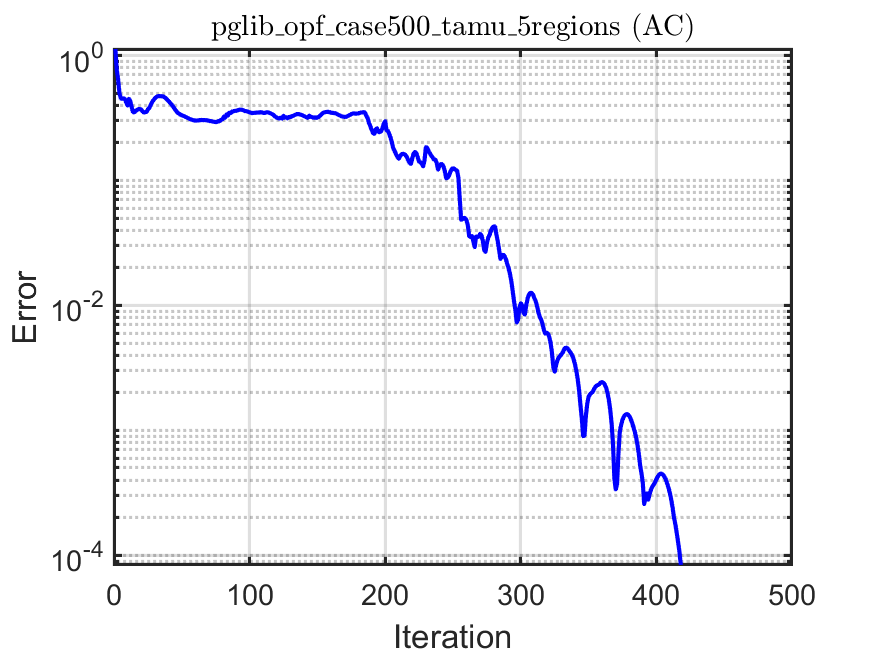}
\includegraphics[width=0.45\linewidth]{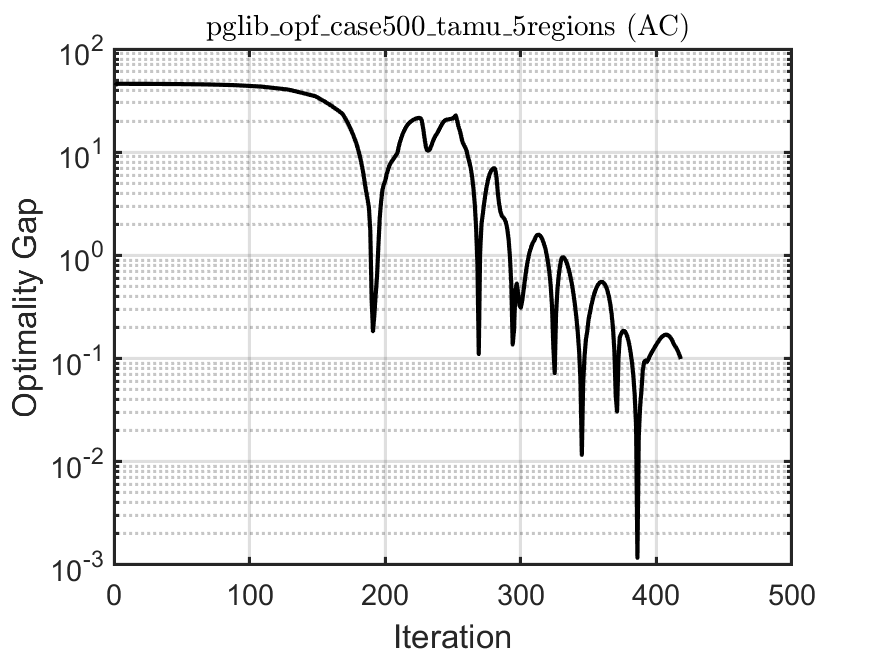}
\caption{primal residual (left) and optimality gap (right)}
\label{fig:case500_ac}
\end{figure}

For \texttt{pglib\_opf\_case500\_tamu}, Figures~\ref{fig:case500_dc} and~\ref{fig:case500_ac} illustrate DC and AC residuals and optimality gaps, respectively.

\subsection*{Case pglib\_opf\_case1354\_pegase}

\begin{figure}[t!]
\captionsetup{font={footnotesize}}
\centering
\includegraphics[width=1\columnwidth]{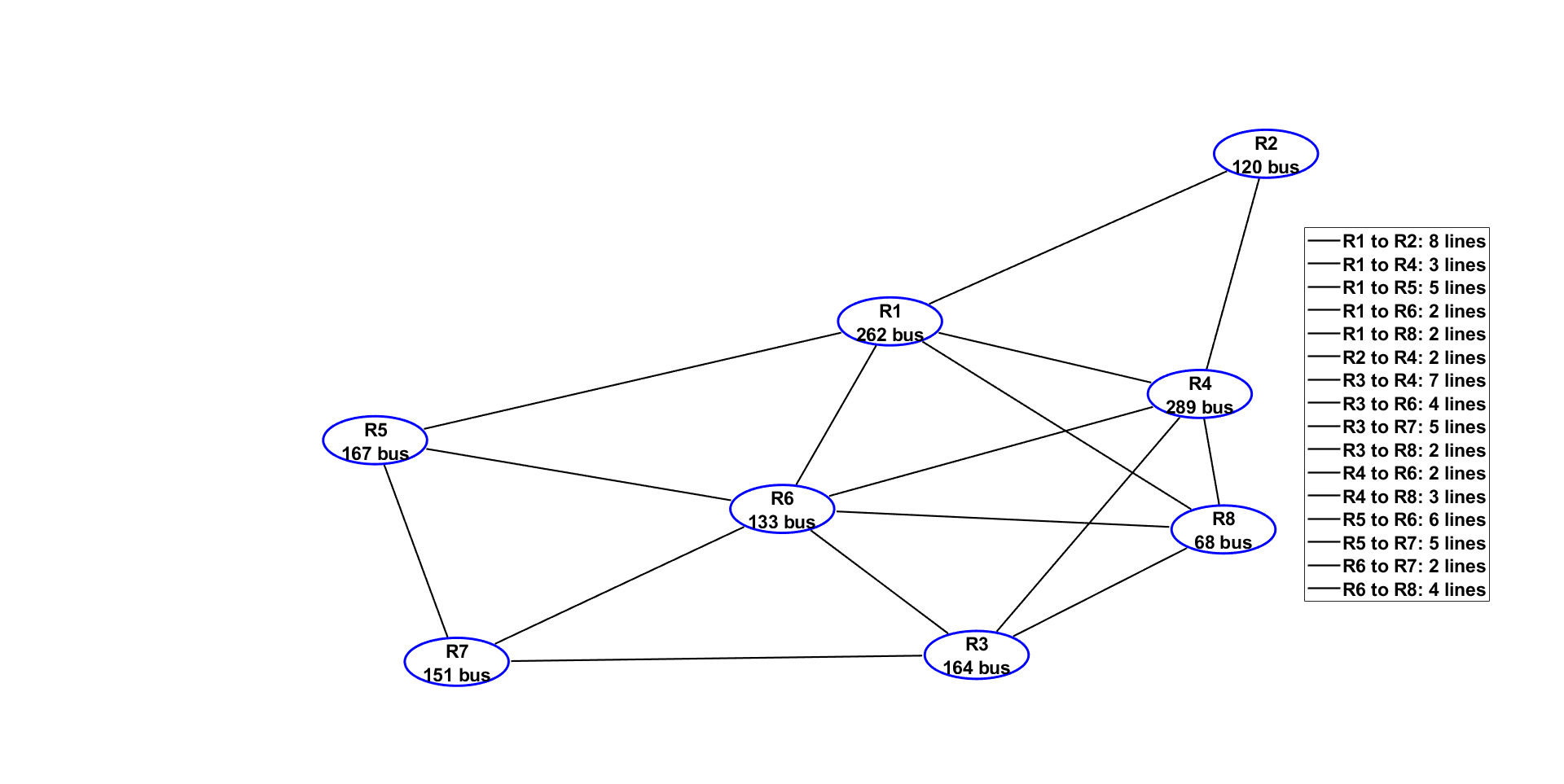}
\caption{Eight-region topology for case1354\_pegase}
\label{fig:topo_1354}
\end{figure}

The dataset \texttt{pglib\_opf\_case1354\_pegase} is part of the PEGASE project, which produced large-scale synthetic European transmission networks intended for stability and optimization studies. This 1354-bus system includes high-density meshed areas characteristic of continental European grids. It is partitioned into eight regions with numerous inter-region connections reflecting the underlying network complexity. The multi-region layout is shown in Fig.~\ref{fig:topo_1354}.

\begin{figure}[t!]
\captionsetup{font={footnotesize}}
\centering
\includegraphics[width=0.45\linewidth]{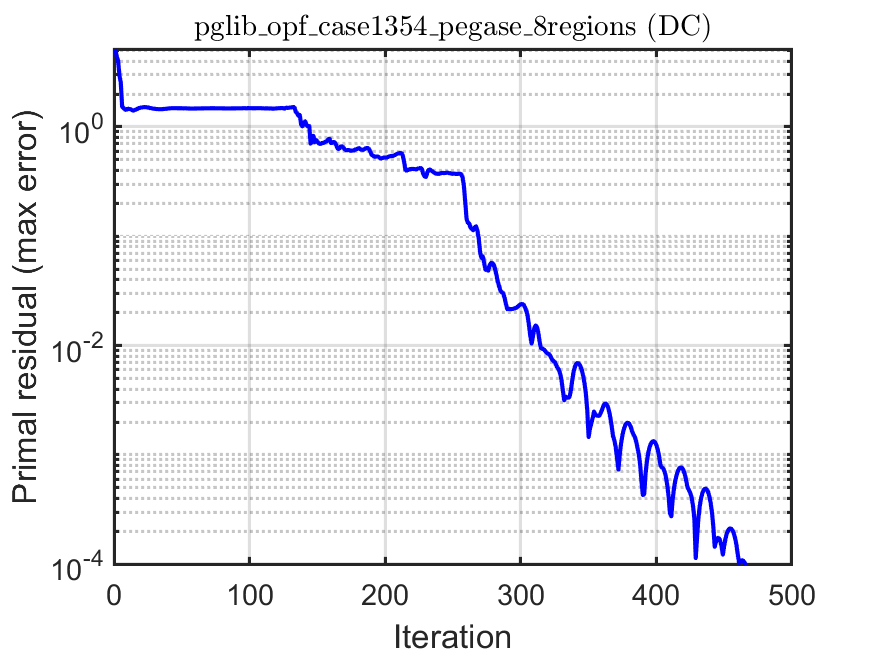}
\includegraphics[width=0.45\linewidth]{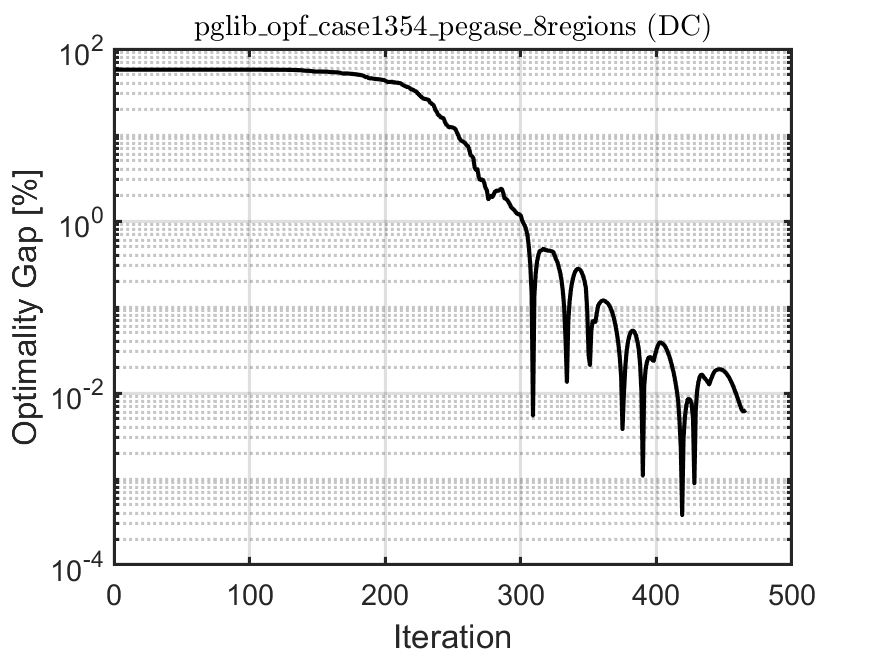}
\caption{primal residual (left) and optimality gap (right)}
\label{fig:case1354_dc}
\end{figure}

\begin{figure}[t!]
\captionsetup{font={footnotesize}}
\centering
\includegraphics[width=0.45\linewidth]{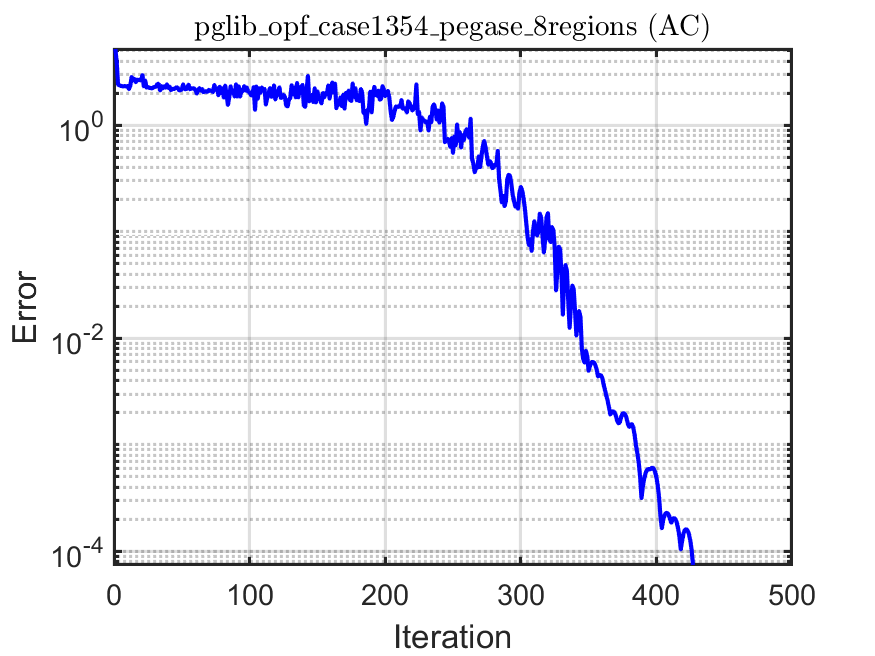}
\includegraphics[width=0.45\linewidth]{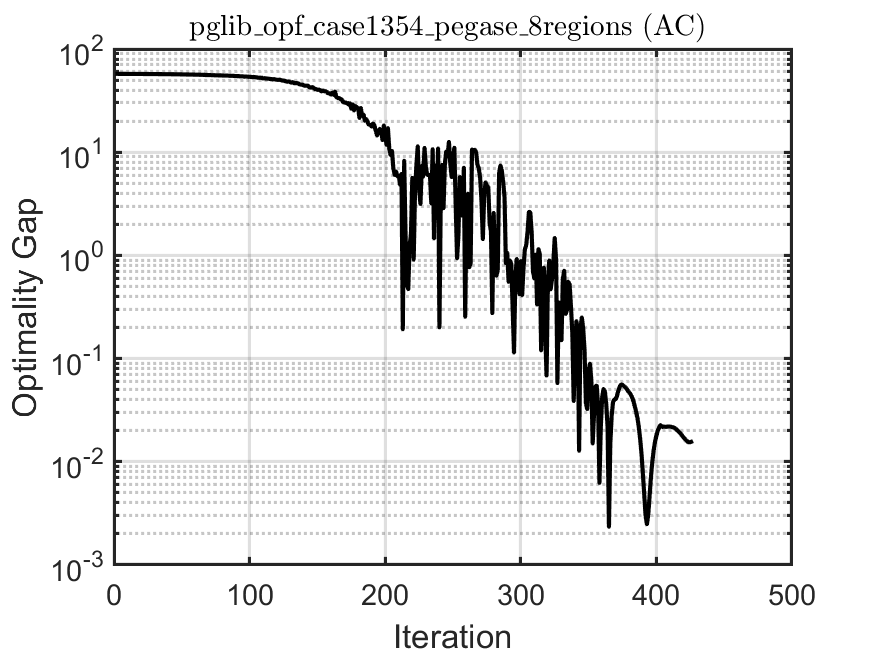}
\caption{primal residual (left) and optimality gap (right)}
\label{fig:case1354_ac}
\end{figure}

Figures~\ref{fig:case1354_dc} and~\ref{fig:case1354_ac} present the DC and AC verification curves for \texttt{pglib\_opf\_case1354\_pegase}. 

\subsection*{Case pglib\_opf\_case2869\_pegase}

\begin{figure}[t!]
\captionsetup{font={footnotesize}}
\centering
\includegraphics[width=1\columnwidth]{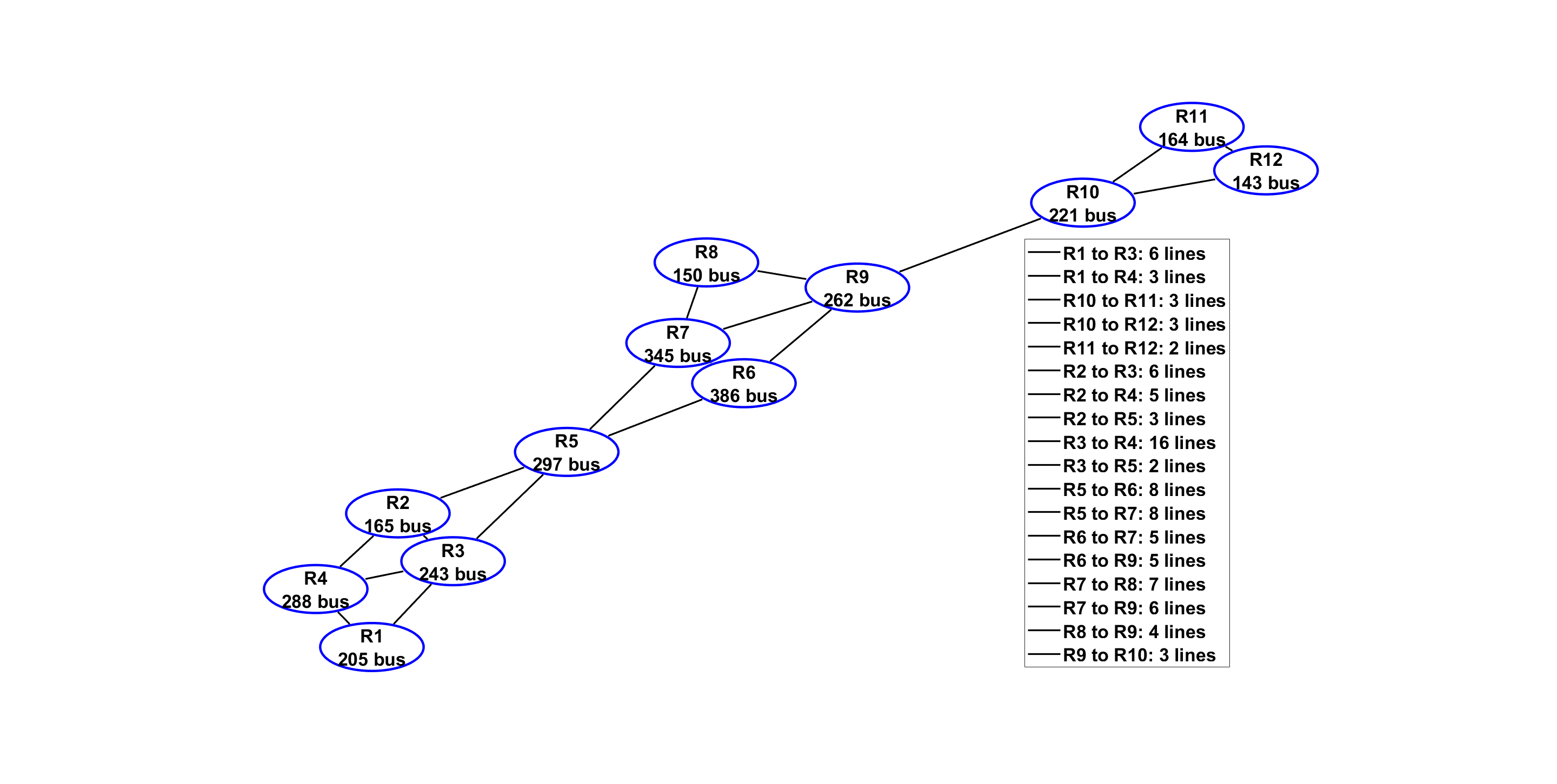}
\caption{Twelve-region topology for case2869\_pegase}
\label{fig:topo_2869}
\end{figure}

The system \texttt{pglib\_opf\_case2869\_pegase} represents a large portion of the European transmission network, consisting of 2869 buses, 510 generators, and 4582 branches. The partitioning procedure divides the system into 12 regions, interconnected by 95 tie-lines. The resulting multi-region topology is shown in Fig.~\ref{fig:topo_2869}. All buses, generators, and branches are fully accounted for across the regions, and the reference bus (Bus 1314) is assigned to Region~6.

\begin{figure}[t!]
\captionsetup{font={footnotesize}}
\centering
\includegraphics[width=0.45\linewidth]{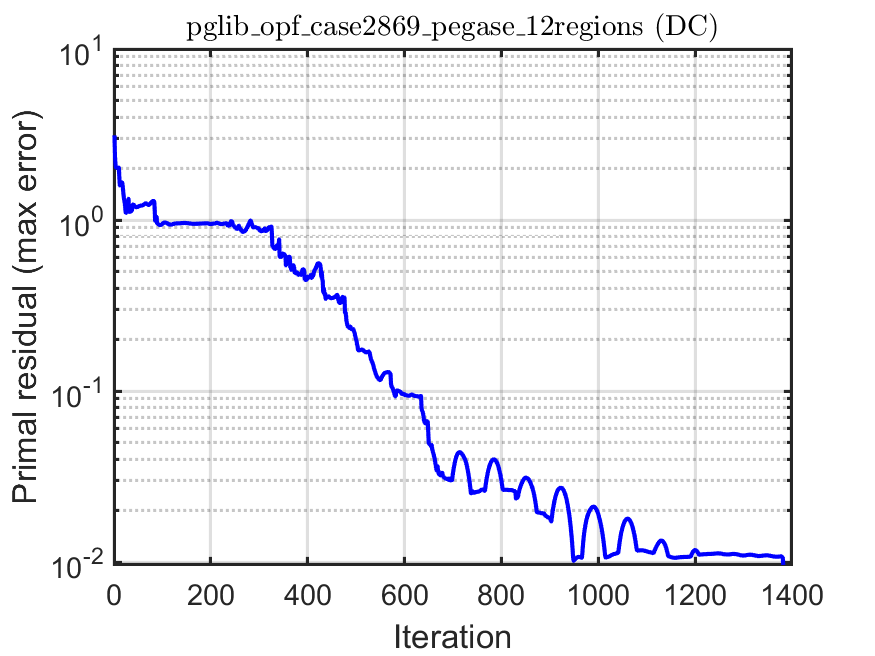}
\includegraphics[width=0.45\linewidth]{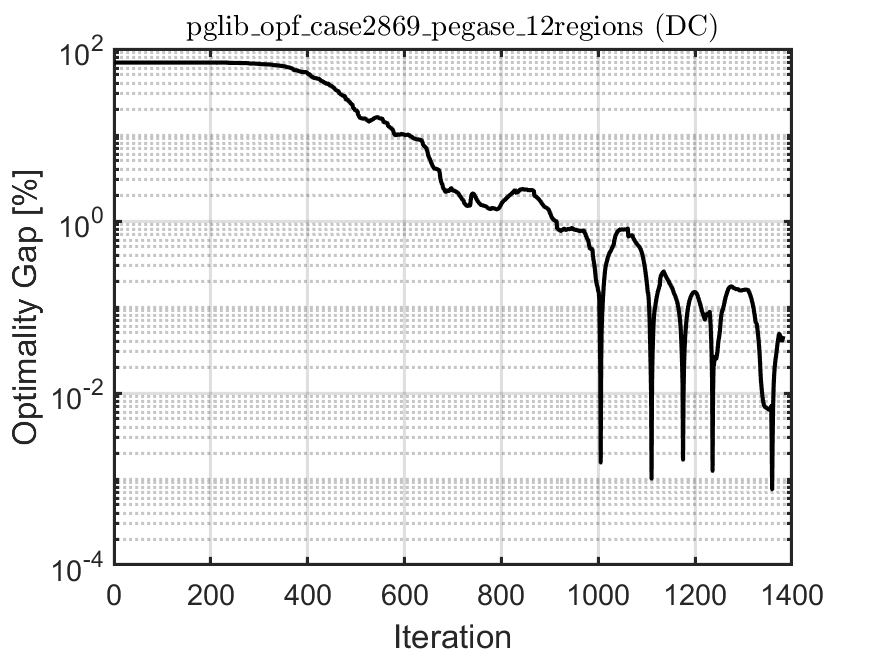}
\caption{Primal residual (left) and optimality gap (right)}
\label{fig:case2869_dc}
\end{figure}

\begin{figure}[t!]
\captionsetup{font={footnotesize}}
\centering
\includegraphics[width=0.45\linewidth]{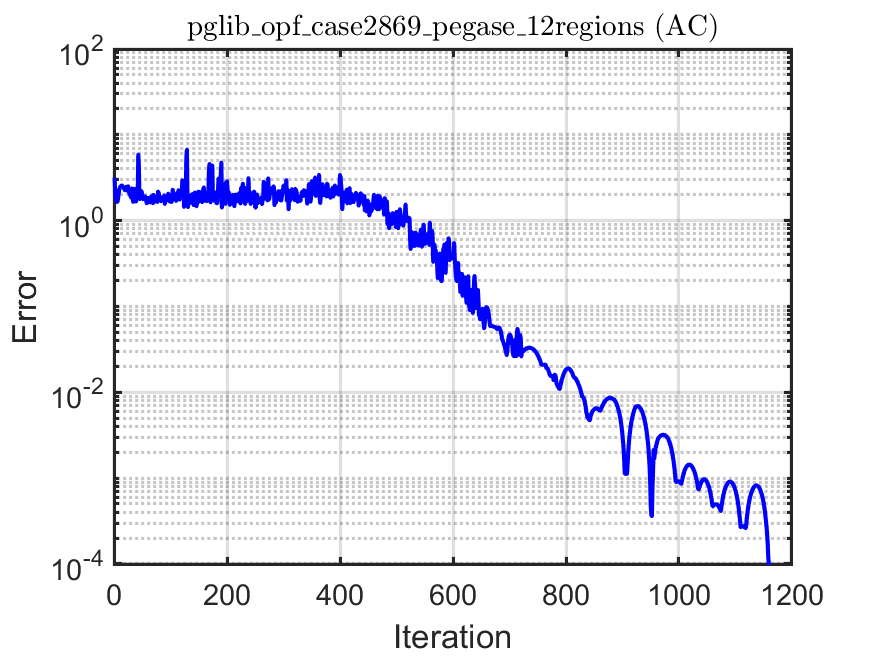}
\includegraphics[width=0.45\linewidth]{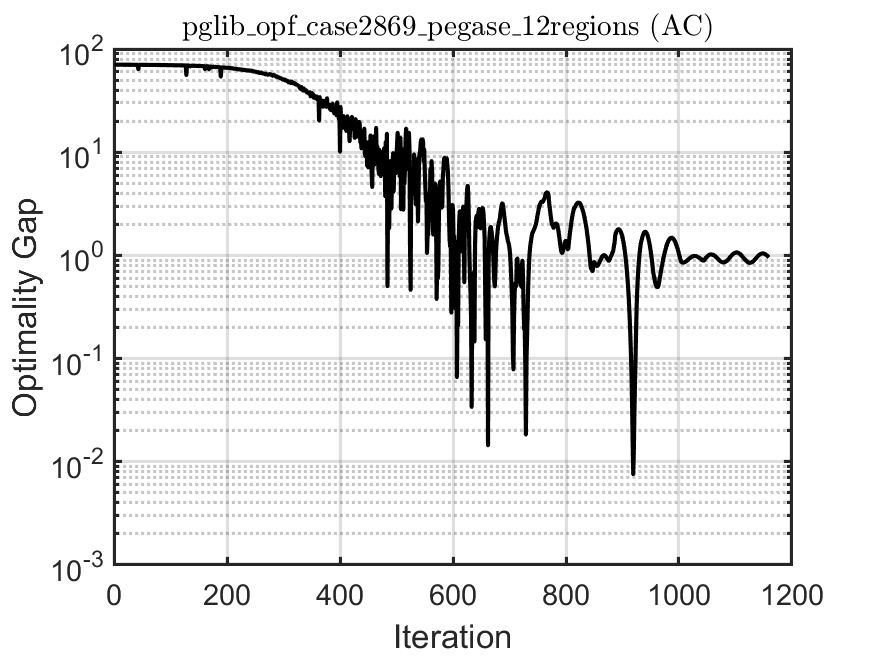}
\caption{Primal residual (left) and optimality gap (right)}
\label{fig:case2869_ac}
\end{figure}

For \texttt{pglib\_opf\_case2869\_pegase}, Figures~\ref{fig:case2869_dc} and~\ref{fig:case2869_ac} illustrate that both DC and AC primal residuals converge to the threshold, while the optimality gaps remain below one percent.

\subsection*{Case pglib\_opf\_case4661\_sdet}

\begin{figure}[t!]
\captionsetup{font={footnotesize}}
\centering
\includegraphics[width=1\columnwidth]{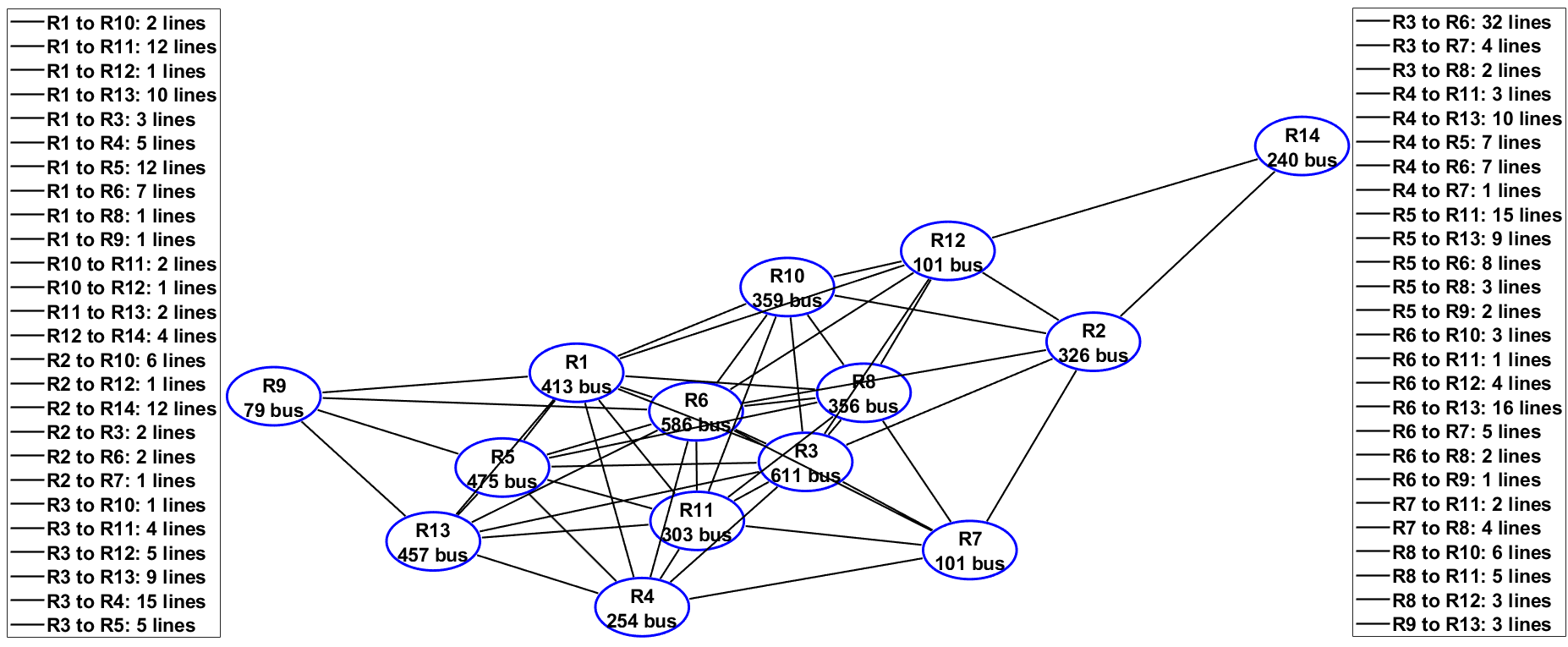}
\caption{Fourteen-region topology for case4661\_sdet}
\label{fig:topo_4661}
\end{figure}

The system \texttt{pglib\_opf\_case4661\_sdet} is a large-scale synthetic model developed using the Sustainable Data Evolution Technology (SDET) framework. With 4661 buses and dense connectivity, it reflects realistic large area transmission dynamics. The system is divided into 14 regions connected by 284 tie-lines. The partitioned structure is shown in Fig.~\ref{fig:topo_4661}.

\begin{figure}[t!]
\captionsetup{font={footnotesize}}
\centering
\includegraphics[width=0.45\linewidth]{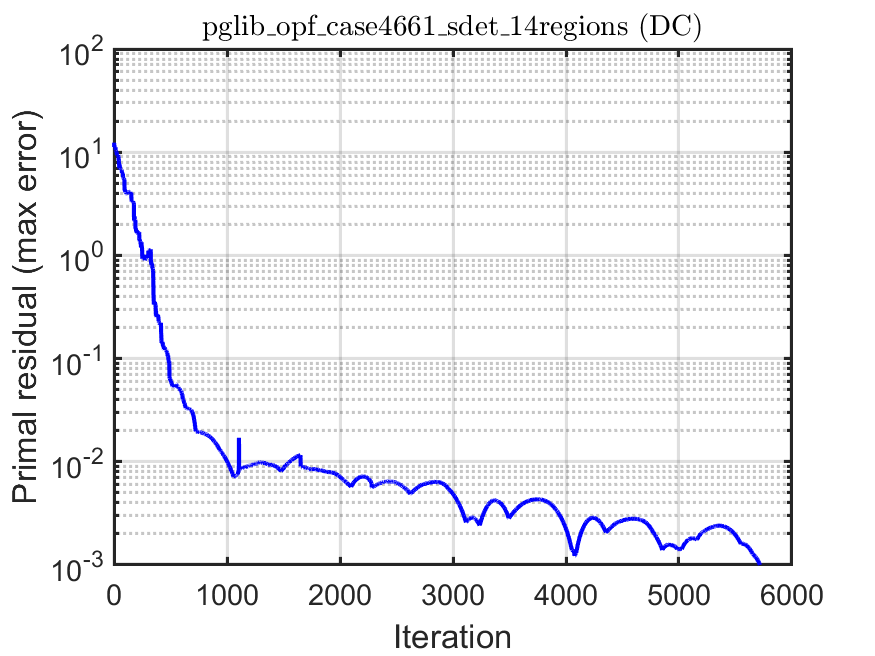}
\includegraphics[width=0.45\linewidth]{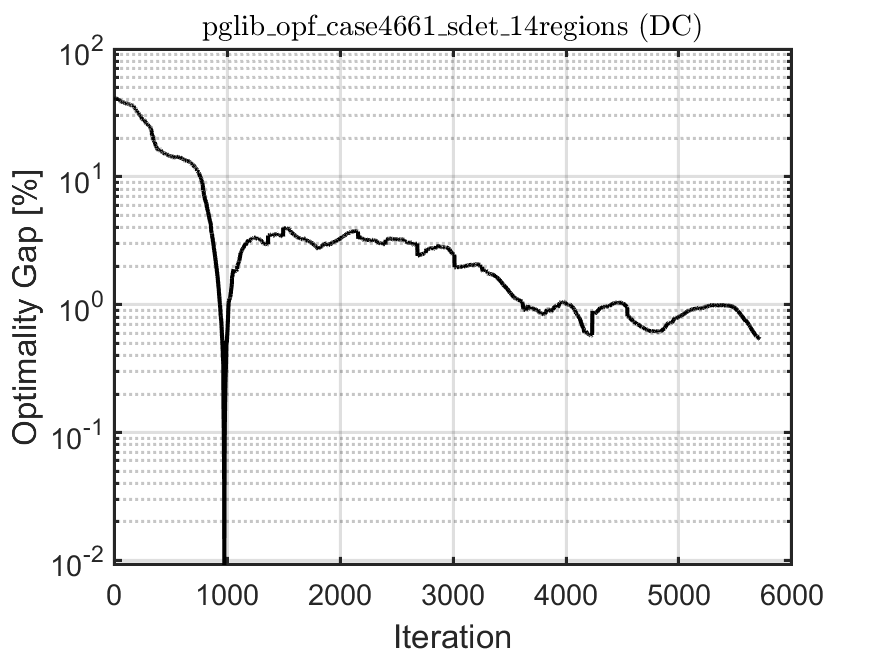}
\caption{primal residual (left) and optimality gap (right)}
\label{fig:case4661_dc}
\end{figure}

\begin{figure}[t!]
\captionsetup{font={footnotesize}}
\centering
\includegraphics[width=0.45\linewidth]{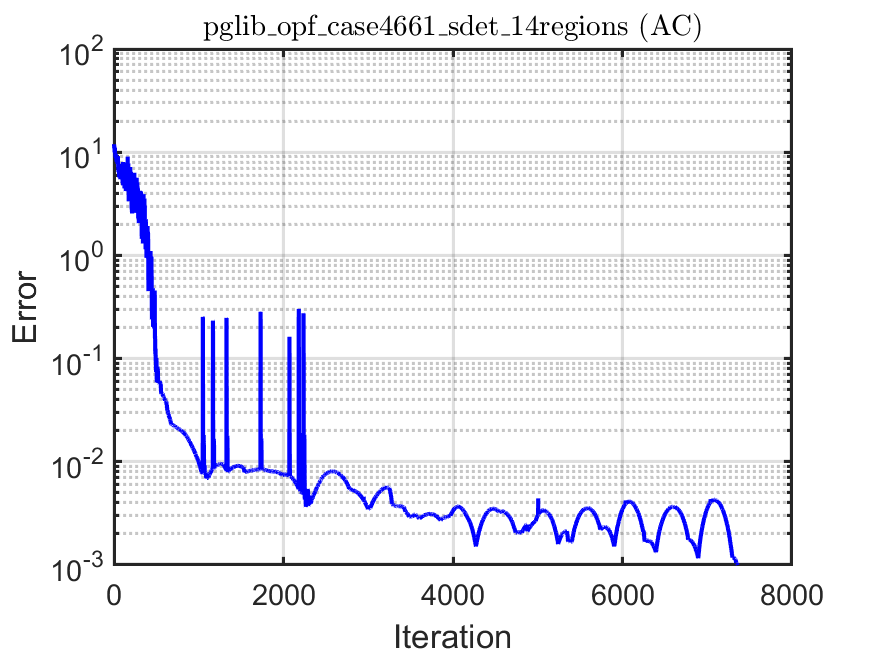}
\includegraphics[width=0.45\linewidth]{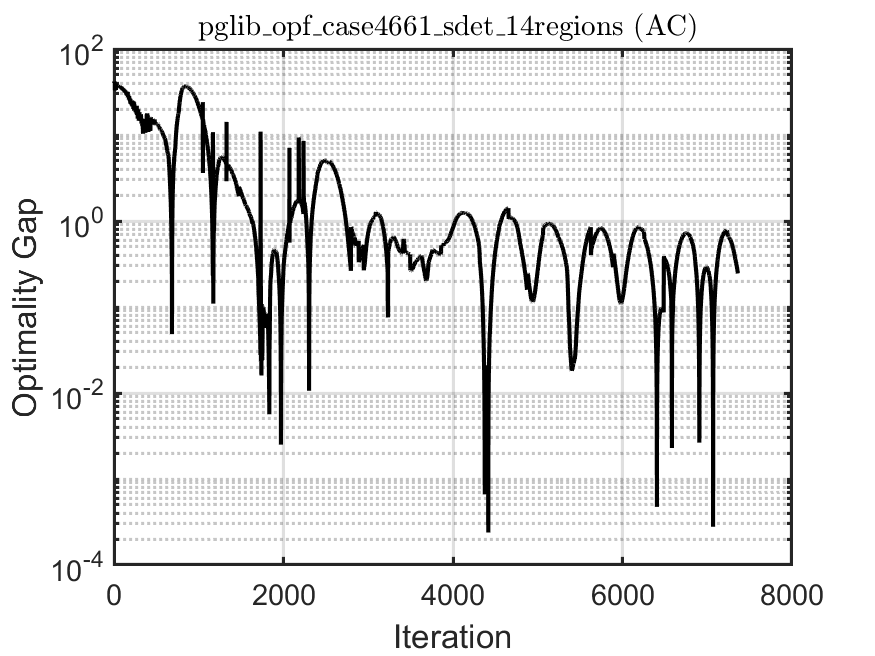}
\caption{primal residual (left) and optimality gap (right)}
\label{fig:case4661_ac}
\end{figure}

For \texttt{pglib\_opf\_case4661\_sdet}, Figures~\ref{fig:case4661_dc} and~\ref{fig:case4661_ac} indicate that the DC and AC primal residuals converge to the threshold, with optimality gaps remaining below one percent.

\subsection*{Case pglib\_opf\_case9241\_pegase}

\begin{figure}[t!]
\captionsetup{font={footnotesize}}
\centering
\includegraphics[width=1\columnwidth]{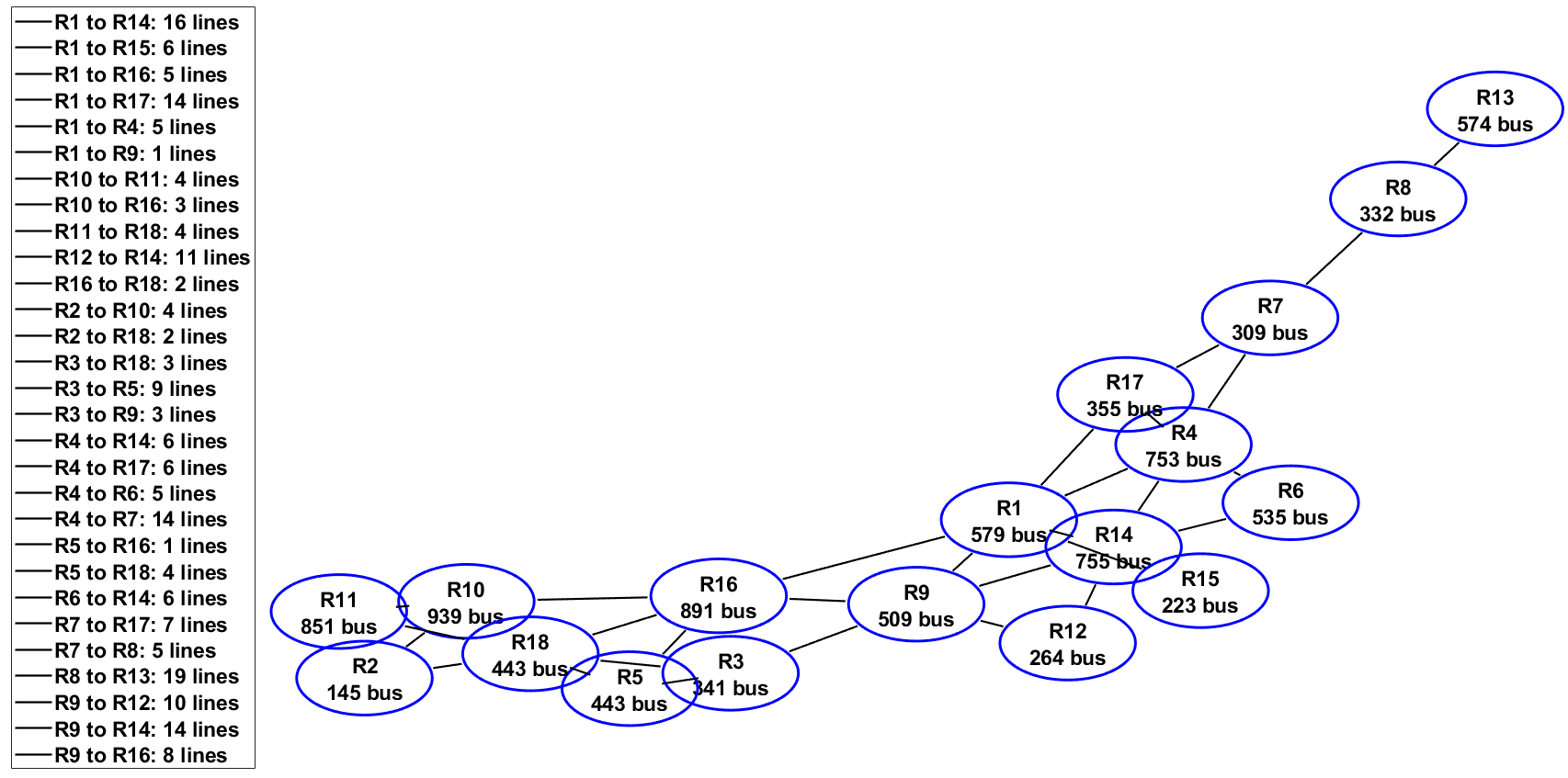}
\caption{Eighteen-region topology for case9241\_pegase}
\label{fig:topo_9241}
\end{figure}

The largest benchmark system considered in DPLib is \texttt{pglib\_opf\_case9241\_pegase}, a synthetic European transmission network derived from the PEGASE project. With 9241 buses and more than sixteen thousand branches, it reflects the characteristics of a highly interconnected continental-scale system. The partitioning procedure yields 18 regions linked by 197 tie-lines. The resulting multi-region topology is shown in Fig.~\ref{fig:topo_9241}.

\begin{figure}[t!]
\captionsetup{font={footnotesize}}
\centering
\includegraphics[width=0.45\linewidth]{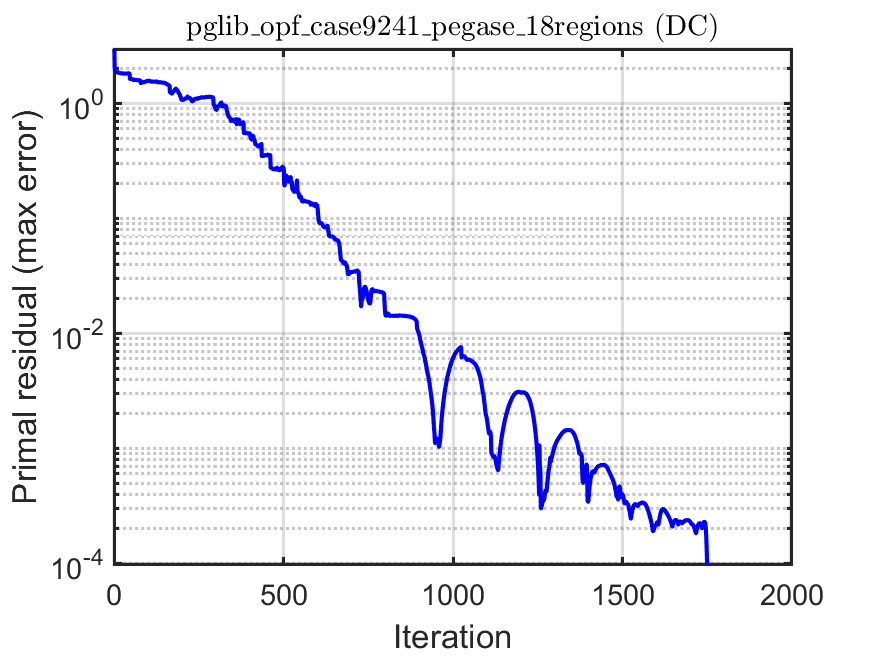}
\includegraphics[width=0.45\linewidth]{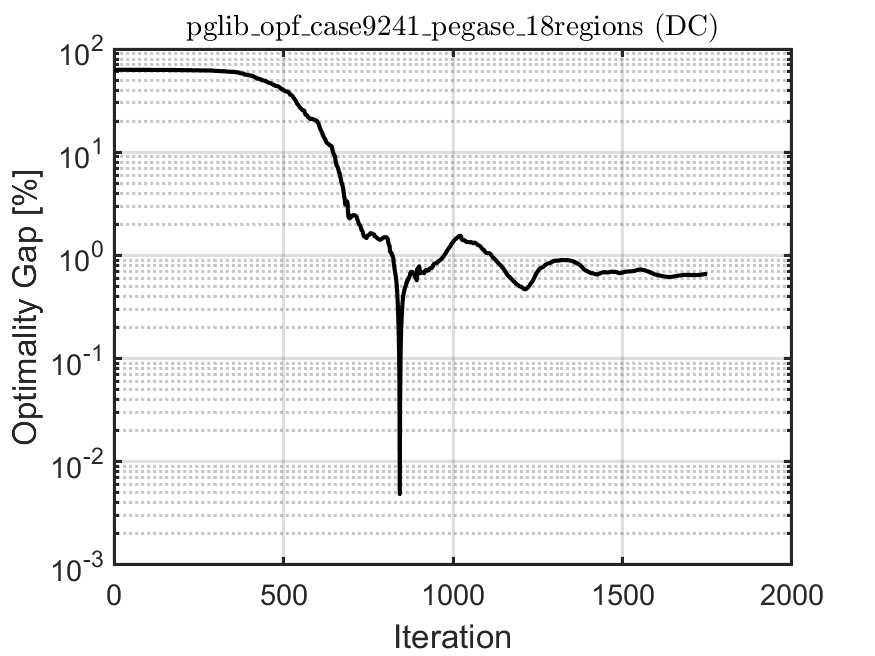}
\caption{primal residual (left) and optimality gap (right)}
\label{fig:case9241_dc}
\end{figure}

\begin{figure}[t!]
\captionsetup{font={footnotesize}}
\centering
\includegraphics[width=0.45\linewidth]{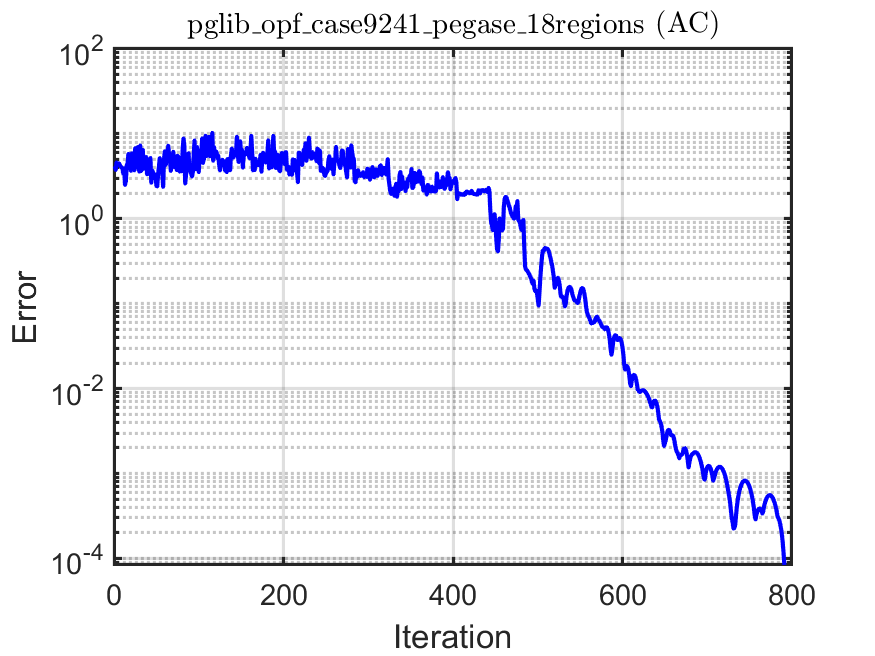}
\includegraphics[width=0.45\linewidth]{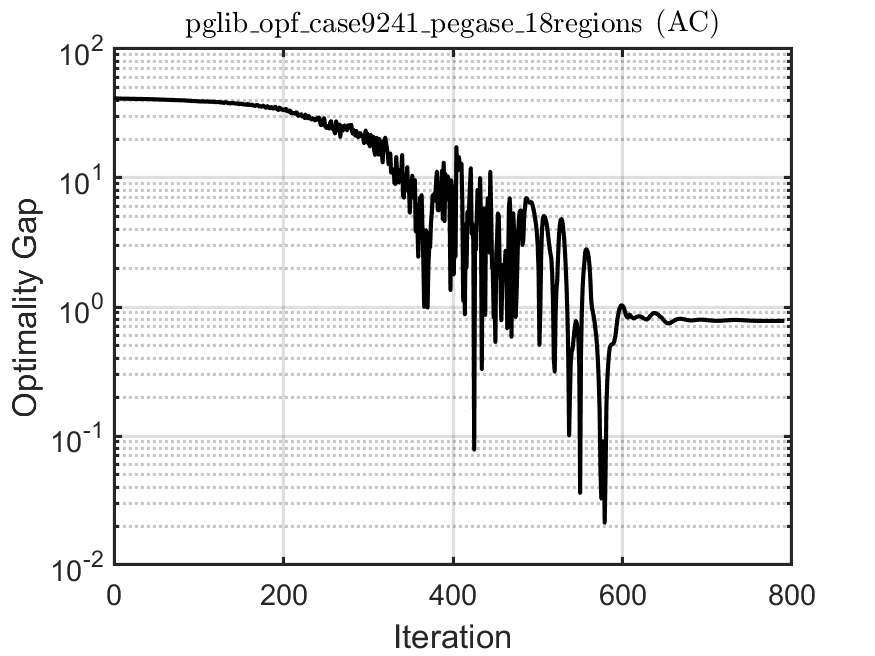}
\caption{primal residual (left) and optimality gap (right)}
\label{fig:case9241_ac}
\end{figure}

For \texttt{pglib\_opf\_case9241\_pegase}, Figures~\ref{fig:case9241_dc} and~\ref{fig:case9241_ac} confirm that the DC and AC primal residuals converge to the threshold, with optimality gaps remaining within one percent, even for this largest benchmark.

{\color{black}

Although distributed OPF is used as the main validation workflow in this paper, the generated DPLib datasets can also support other distributed power system applications. As an illustrative example beyond OPF, we use DPLib to prepare regional contingency screening data for \texttt{pglib\_opf\_case118\_ieee\_3regions}. In this workflow, each candidate outage is mapped to the regional dataset affected by that outage. An internal branch outage modifies only the regional MATPOWER-compatible file containing that branch. A tie-line outage affects the two neighboring regions connected by the tie-line and the corresponding inter-regional tie-line record. A generator outage is assigned to the region containing the affected generator. Therefore, the DPLib regional cases, local/global bus mappings, generator mappings, and explicit tie-line records provide the information needed to construct screening-ready contingency datasets without manually tracing the centralized to regional conversion.
}

\begin{table}[!t]
\centering
\captionsetup{font={footnotesize}}
{\color{black}
\caption{Regional contingency screening data}
\label{tab:contingency_screening_case118}
\scriptsize
\setlength{\tabcolsep}{4pt}
\renewcommand{\arraystretch}{1.0}
\begin{tabular}{lrrr}
\toprule
\textbf{Region} & \textbf{Buses} & \textbf{Internal Branch Cont.} & \textbf{Gen. Cont.} \\
\midrule
R1 & 35 & 55 & 17 \\
R2 & 45 & 70 & 20 \\
R3 & 38 & 51 & 17 \\
\midrule
Total internal & 118 & 176 & 54 \\
tie-line cont. & -- & 10 & -- \\
Total branch cont. & -- & 186 & -- \\
\bottomrule
\end{tabular}
}
\end{table}

Table~\ref{tab:contingency_screening_case118} reports the resulting contingency screening data summary. The 186 branch contingencies of the centralized case are separated into 176 internal regional branch outages and 10 inter-regional tie-line outages. The 54 generator contingencies are mapped to their corresponding regions. This example is a data preparation workflow rather than a new contingency analysis algorithm. The resulting contingency maps can be used by distributed power flow, distributed security assessment, distributed corrective OPF, reliability screening, or other multi-area studies. More broadly, the same DPLib data structure can support applications such as multi-region unit commitment, where regional generation and tie-line limits are needed; transmission and generation expansion planning, where candidate investments can be assigned to regions and inter-regional corridors; and voltage stability analysis, where regional buses, boundary buses, and tie-line exchanges must be tracked consistently. This demonstrates that DPLib is not limited to OPF validation; rather, it provides a reusable distributed data layer for broader power system analysis.

\section{Conclusion}
\label{sec:conclusion}

This work presented DPLib, a reproducible, open-source benchmark library for distributed power system studies. DPLib addresses the gap in standardized distributed datasets by offering a graph-based partitioning toolkit and 40 multi-region test cases ranging from 5 buses to 20758 buses, derived from MATPOWER-compatible systems. The generated datasets provide complete regional MATPOWER-compatible cases with explicit tie-line and mapping information, and the toolkit supports both automatic and user-defined regional partitions. The library is complemented by modular ADMM-based DC and AC OPF solvers, which validate generated regional datasets across diverse case sizes.

Numerical results compare the DPLib partitioning options with METIS, KaFFPa, and an IPA-inspired baseline, showing tradeoffs among tie-line count, regional bus balance, and complete distributed data generation. The validation results report centralized run times, distributed OPF iteration counts, run times, and optimality gaps, demonstrating that the generated datasets can support complete distributed optimization workflows. A regional contingency screening example also demonstrates DPLib's use beyond OPF.

To support long-term usability, DPLib will continue to be maintained as an actively developed open source project. The repository will remain versioned through GitHub, where users may report issues, request features, and contribute to the code.

	\bibliographystyle{IEEEtran}
	\bibliography{example}
	
\end{document}